\documentclass[%
  article,notitlepage,
 reprint,
 amsmath,amssymb,
 aps,
 floats,
]{revtex4-1}

\usepackage{graphicx}% Include figure files
\usepackage{dcolumn}% Align table columns on decimal point
\usepackage{bm}% bold math
\usepackage{color}
\usepackage{amssymb}

\begin{document}

\title{Diffusiophoretically induced interactions between chemically active and
  inert particles}
\thanks{Electronic Supplementary Information (ESI) available:
    Two movies of two spheres motions, the catalytic sphere fixed in space ($S1$)
    and both spheres freely moving ($S_2$).}
  
\author{Shang Yik Reigh}
\email{reigh@is.mpg.de}
\affiliation{Max-Planck-Institut f{\"u}r Intelligente Systeme,
  Heisenbergstra{\ss}e 3, 70569 Stuttgart, Germany}
\author{Prabha Chuphal}
\email{prabhac@iiserb.ac.in}
\affiliation{Department of Physics, Indian Institute of
    Science Education and Research Bhopal, India}
\author{Snigdha Thakur}
\email{sthakur@iiserb.ac.in}
\affiliation{Department of Physics, Indian Institute of
    Science Education and Research Bhopal, India}
\author{Raymond Kapral}
\email{rkapral@chem.utoronto.ca}
\affiliation{Chemical Physics Theory Group, Department of
    Chemistry, University of Toronto, Toronto, Ontario M5S 3H6}

\date{\today}

\begin{abstract}
  In the presence of a chemically active particle, a nearby chemically inert particle
  can respond to a concentration gradient and move by diffusiophoresis. The nature
  of the motion is studied for two cases: first, a fixed reactive
  sphere and a moving inert sphere, and second, freely moving reactive and inert spheres.
  The continuum reaction-diffusion and Stokes equations are solved analytically for these systems and
  microscopic simulations of the dynamics are carried out. Although the relative velocities of the spheres are very similar in the two systems, the local and global structures of streamlines and the flow velocity fields are found to be quite different. For freely moving spheres, when the two spheres approach each other the flow generated by the inert sphere through diffusiophoresis drags the reactive sphere towards it. This leads to a
  self-assembled dimer motor that is able to propel itself in solution.
  The fluid flow field at the moment of dimer formation changes direction. The
  ratio of sphere sizes in the dimer influences the characteristics of the
  flow fields, and this feature suggests that active self-assembly of
  spherical colloidal particles may be manipulated by sphere-size changes in
  such reactive systems.
\end{abstract}

\pacs{Valid PACS appear here}

\maketitle

\section{Introduction}

Both living organisms and inanimate objects can respond to the presence of
chemical gradients by moving either towards or away from high concentrations
of chemical species. In the biological realm organisms are observed to orient or move in response to chemical agents. For instance, \textit{E. coli} bacteria are found in glucose-rich regions indicating that they search for food and tend to migrate toward it~\cite{berg72,berg04}, sperm cells are known to follow concentration gradients of chemoattractants secreted by the oocyte for fertilization~\cite{eisenbach:06}, and there are many other examples.~\cite{berg72,adler75}
The ability to sense chemical gradients is not restricted to living
organisms. It is well known that colloidal particles can respond to chemical
gradients and move to higher or lower concentration regions through
diffusiophoretic mechanisms~\cite{derjaguin:47,derjaguin:74,anderson86,anderson:89}.
In this and other phoretic mechanisms, the gradient of some field across the colloidal particle gives rise to a body force, which, because of momentum conservation, induces fluid flow in the surrounding medium that causes the particle to move.
The motions of motors propelled by self-phoretic
mechanisms~\cite{golestanian:05,howse:07,kapral:13,colberg14} have also been observed
to be affected by the presence of chemical gradients; for example, experiments
have shown that bimetallic-rod and Janus motors preferentially move towards
higher hydrogen peroxide concentrations.~\cite{hong:07,baraban:13} As well,
simulations of sphere-dimer motors in a microfluidic channel~\cite{deprez:17}
and in bulk solution~\cite{chen:16} show that these motors respond to
concentration gradients.

In this article, we investigate the dynamics of a pair of small colloidal particles, one of which is chemically active and converts fuel to product, while the other is nonreactive. Further, we suppose that the interactions of the fuel and product molecules with the colloidal particles are the same for the reactive particle but different for the nonreactive particle, so that the nonreactive particle can respond to the chemical gradient produced by the catalytic particle as a result of diffusiophoresis. We consider interactions such that diffusiophoresis causes motion towards high product concentrations, and situations where the reactive particle is either fixed or free to move.

These specific choices are only a few among several other
possibilities. For instance, the interaction potentials may be chosen so
that either or both colloidal particles may be diffusiophoretically active
with different responses to gradients.~\cite{footnote:back} Also, either
particle may be fixed or free to move, or their internuclear separation can
be fixed as in a sphere-dimer
motor~\cite{ruckner:07,ozin:10,reigh:15dimer}. 
All of these situations
are potentially interesting to study. A study, based on a continuum
description of the fluid, of the dynamics of a pair of colloidal particles
each of which could be Janus particles or active or inert is related to the
work presented here.~\cite{sharifi:16,najafi:16} In order to
investigate the dynamical properties of the spheres we use deterministic
continuum theory as well as coarse-grain microscopic simulations. The
particle-based simulations include fluctuations relevant for experimental
studies of small active colloidal particles in solution,~\cite{fischer:14}
and automatically account for chemical-gradient, hydrodynamic and direct intermolecular interactions between the spheres without imposing specific boundary conditions.~\cite{footnote:mpc1}

The diffusiophoretic mechanism for the motion of a colloidal particle in
an external concentration gradient is well
known.~\cite{derjaguin:47,derjaguin:74,anderson86,anderson:89} By choosing
the fixed reactive particle in our study to be diffusiophoretically
inactive, it serves simply as reactive source that produces concentration
gradients in the system.~\cite{hong:07,baraban:13,chen:16,deprez:17} The nonreactive
colloidal particle responds to this chemical gradient, which is analogous to
an external chemical gradient, but presents some additional features as a
result of pinning and reaction. We may contrast this case with that when the
reactive sphere is free to move. The reactive particle again only generates concentration gradients in the system but when the two spheres closely approach we show that they form a self-assembled sphere-dimer motor that moves autonomously in solution, and we find that substantial changes in the flow fields occur at the moment of the dimer formation.

On a basic level, investigations of the mechanisms that give rise to the
concentration and fluid flow fields that are responsible for the dynamics
provide insight into the relative roles of chemical and hydrodynamic
interactions, a topic that is important for studies of the collective dynamics
of active particles.~\cite{palacci:13,buttinoni:13,singh:17} In this
connection, recent experimental and computational studies have considered
mixtures of chemically active and inactive spherical particles that exhibit
interesting self-assembly and emergent
dynamics.~\cite{balazs:17,schmidt:18,Yu:18} As in the present study, the
dynamics of such mixtures will depend on both hydrodynamic and chemical,
temperature, or electric fields that exist in the system.~\cite{keh:89,ishikawa06,yang:14,najafi:16,sharifi:16,niu:17,ni:17}

In Sec.~\ref{sec:cont-theory} we present continuum solutions for the reaction-diffusion and Stokes equations for this problem, and Sec.~\ref{sec:sim} describes the particle-based simulation method. Sections~\ref{sec:fix} and \ref{sec:move} discuss the physical phenomena that are observed for fixed and freely moving catalytic spheres, respectively. The conclusions of the investigation are given in Sec.~\ref{sec:conc}.

\section{Continuum theory}\label{sec:cont-theory}
%\subsection{Model description}
\begin{figure}[t]
  \centering
  \includegraphics[scale=0.6]{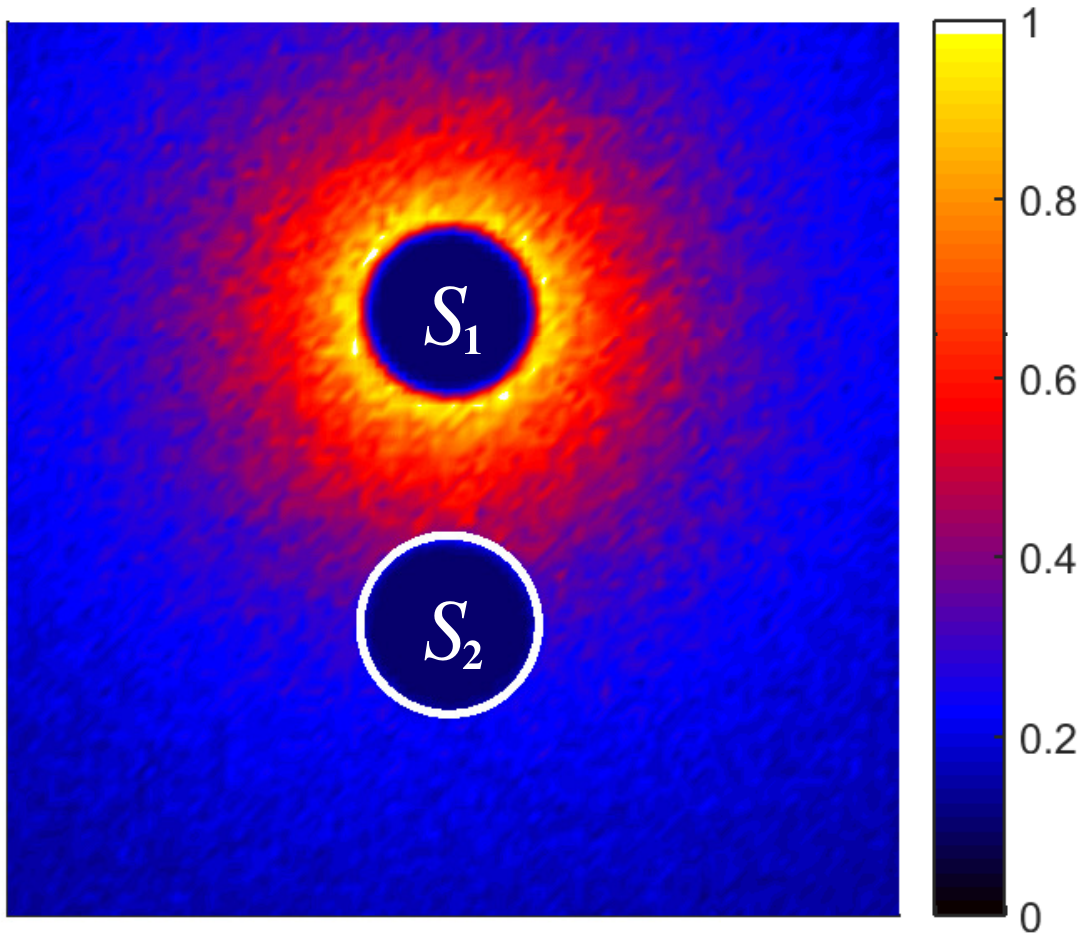}
  \caption{
    Two spheres, one catalytically active ($S_1$) and the other catalytically inactive ($S_2$), are shown.
    The $S_1$ sphere, as a source of concentration gradients, converts species $A$ (reactant) to $B$ (product) in the
    reaction, $A + S_1 \rightarrow B + S_1$, which generates
    inhomogeneous concentration fields around the $S_2$ sphere.
    The $S_2$ sphere moves by the diffusiophoretic mechanisms due to the
    asymmetry of the concentration field in its vicinity.
    The numbers in the color bar indicate the normalized concentration
    of products ($B$).
    (The figure is constructed from simulation data described in the text. The sphere separation distance is $L/\sigma=3.5$.)
}
  \label{fig:intro}
\end{figure}
\begin{figure}%[htbp]
  \centering
  \includegraphics[scale=0.7]{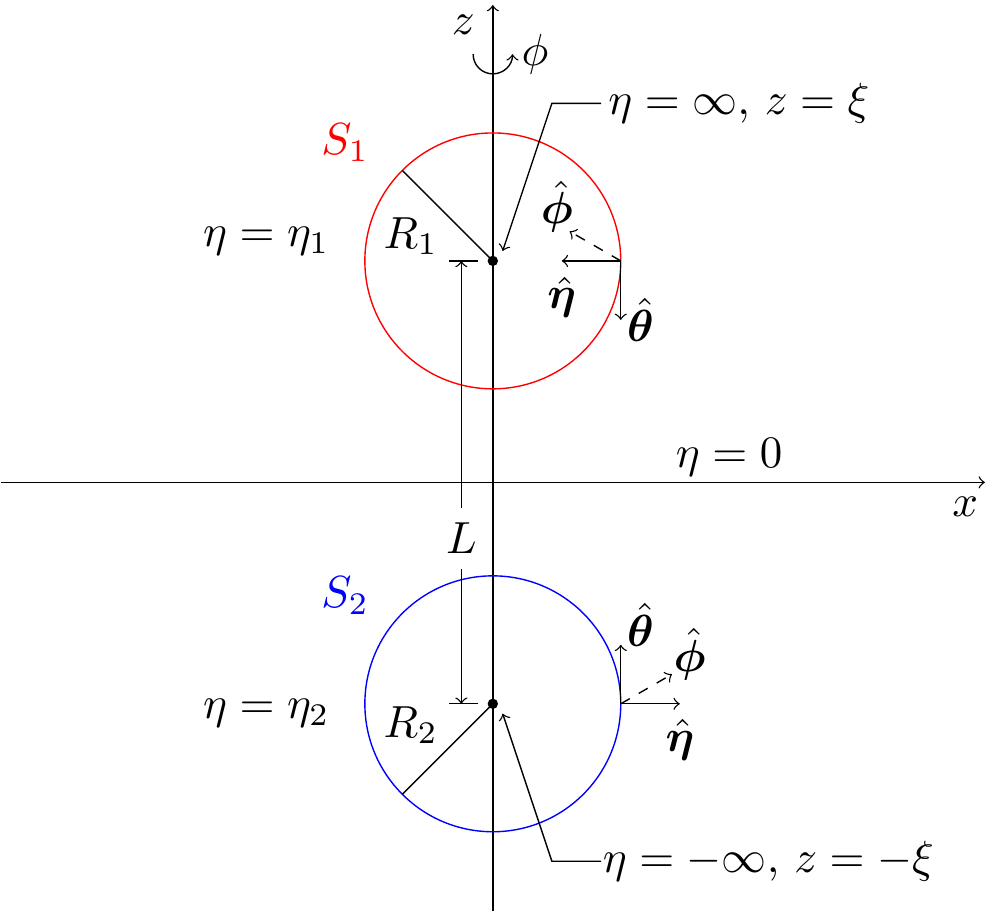}
  \caption{Bispherical ($\theta$, $\eta$, $\phi$) and Cartesian ($x$, $y$, $z$) coordinates for two spheres. The catalytic sphere $S_1$ (red) with radius $R_1$ and noncatalytic sphere $S_2$ (blue) with radius $R_2$, separated by a distance $L$, can be specified by variables $\eta=\eta_1$ and $\eta=\eta_2$, respectively. The system is axisymmetric in the angle $\phi$ about the $z$ axis that lies along the line connecting the centers of the two spheres. The hat notation is used to indicate unit vectors.
  }
  \label{fig:coord}
\end{figure}

We consider two spheres, a catalytically active sphere $S_1$ with radius $R_1$
and a catalytically inactive sphere $S_2$ with radius $R_2$. These spheres,
shown in Fig.~\ref{fig:intro}, are taken to be separated by a distance $L$ in
three dimensional space. Two solute species $A$ (reactant) and $B$ (product)
take part in the irreversible chemical reaction $A + S_1 \rightarrow B + S_1$
on the catalytic sphere. Since we consider the case where catalytic sphere has
no phoretic mobility, the interaction potentials of these species with the
catalytic sphere are assumed to be the same, $U_{1, A}=U_{1, B}$, while they
are different for the noncatalytic sphere, $U_{2, A}\ne U_{2, B}$,
where $U_{i,I}$ is the interaction potential between the sphere $i$ and the
solute species $I$.

In this circumstance the concentration gradient in the system arising from
chemical activity on $S_1$ will induce  a body force on the noncatalytic
sphere $S_2$. The diffusiophoretic mechanism will then operate and lead to a mean velocity component along the line of centers between the two spheres due to the axial symmetry of the system. In the continuum description our interest is in the value of the mean velocity that results from this mechanism, as well as the forms of the concentration and fluid velocity fields that accompany it.

The two-sphere system can be solved in a bispherical coordinate system.~\cite{jeff26,pope11,michelin:15,reigh:15dimer,alexander:17} The bispherical coordinates are ($\theta, \eta, \phi$), where $0\leq\theta\leq\pi$, $-\infty\leq\eta\leq\infty$, and $0\leq \phi\leq2\pi$ as shown in Fig.~\ref{fig:coord}. In Cartesian coordinates ($x, y, z$), the relations, $x = \xi\sin\theta \cos\phi/(\cosh\eta-\cos\theta)$, $y = \xi\sin\theta \sin\phi/(\cosh\eta-\cos\theta)$ and
$z = \xi\sinh\eta/(\cosh\eta-\cos\theta)$ are satisfied with a scale factor $\xi\; (>0)$.~\cite{happel73}
The surfaces of the $S_1$ and $S_2$ spheres are represented by the parameters $\eta=\eta_1 (>0)$ and $\eta=\eta_2 (<0)$, respectively. Inversely, from the values of the radii of the $S_1$ and $S_2$ spheres, $R_1$ and $R_2$, and any separation distance, $L$, which is greater than the sum of their radii, the bispherical coordinate parameters,
$\xi$, $\eta_1$ and $\eta_2$ are found by $\xi=\sqrt{(L^2-R_1^2-R_2^2)^2 -4R_1^2 R_2^2}/2L$, $\eta_1= \ln \{\xi/R_1 +\sqrt{1+(\xi/R_1)^2} \}$, and $\eta_2= -\ln \{\xi/R_2 +\sqrt{1+ (\xi/R_2)^2} \}$.

\subsection{\label{sec:con_field}Concentration field}
We assume the P\'{e}clet number is small so that fluid advection may be neglected and the steady-state concentration field of species $A$, $c_A$, can be found from the solution of the diffusion equation,
\begin{eqnarray}
  \nabla^2 c_A=0,
  \label{diff}
\end{eqnarray}
subject to the radiation and reflecting boundary conditions,
\begin{eqnarray}
  (\bm{J}\cdot \hat{\bm{\eta}})_{\eta=\eta_1} &=&\bar{k}_0 c_A(\eta=\eta_1),\nonumber \\
  (\bm{J}\cdot \hat{\bm{\eta}})_{\eta=\eta_2} &=&0,
  \label{bound}
\end{eqnarray}
on the $S_1$ and $S_2$ spheres, respectively. Here $\bm{J}=-D\nabla c_A$ is
the diffusion flux of species $A$, $D$ is the common diffusion constant of $A$
and $B$, and $\bar{k}_0=k_0/(4\pi R_1^2)$, where $k_0$ is the intrinsic
reaction rate coefficient. There are only $A$ particles infinitely far from
the spheres so that $c_A({r \rightarrow \infty}) = c_0$.

The total concentration $c_0 =c_A + c_B$ is conserved in the reaction-diffusion system with the boundary conditions on the surfaces of the spheres and infinity, and we can write $c_A =c_0 - c_B$ locally; thus, we can eliminate $c_A$ and consider only $c_B$. In bispherical coordinates, the concentration of $B$ is now given by
\begin{align}
  &c_B(\theta,\eta)
  =-\sqrt{\cosh\eta-\mu}\sum_{n=0}^{\infty}[A_n e^{(n+\frac{1}{2})\eta}  \nonumber\\
  &\hspace{120pt} 
  +B_n e^{-(n+\frac{1}{2})\eta}]P_n(\mu),
  \label{sol}
\end{align}
where $P_n(\mu)$ is a Legendre function and $\mu=\cos{\theta}$.
The $A_n$ and $B_n$ coefficients may obtained by following the same procedure
used to obtain the solution for sphere dimers.\cite{reigh:15dimer}
%The local concentration of species $A$ may then be obtained from the conservation condition by $c_A =c_0 - c_B$.

\subsection{Particle velocity, streamlines and flow field}

We examine two situations, the first where the catalytic sphere is fixed in space by an external force and the second where it free to move and the system is force-free. Different velocity fields arise in these cases and give rise to dynamics corresponding to physically different phenomena.

\subsubsection{fixed catalytic sphere}

We suppose that the catalytic sphere $S_1$ is fixed in space by external force and
the noncatalytic sphere $S_2$ is able to move in the solution.
The concentration field around the $S_1$ is asymmetric as given
by Eq.~(\ref{sol}); hence, a flow is generated at the surface of the $S_2$ sphere by the diffusiophoretic mechanism.~\cite{anderson86,anderson:89} The slip velocity is the fluid velocity at the outer edge of a boundary layer beyond which the interaction potentials vanish, and is given in the
body-fixed frame of the sphere by
\begin{align}
  \bm{v}_s = -\kappa (\bm{I}-\hat{\bm{n}}\hat{\bm{n}}) \cdot \nabla c_B,
  \label{slip_vel}
\end{align}
where $\bm{I}$ is the unit dyadic, $\hat{\bm{n}}$ the surface normal vector,
\begin{equation}\label{eq:kappa}
\kappa = \frac{k_BT}{ \bar{\mu}} \int_0^\infty
r[e^{-U_{2,B}(r)/(k_BT)}-e^{-U_{2,A}(r)/(k_BT)}]dr,
\end{equation}
is the diffusiophoretic factor, with $\bar{\mu}$ the shear viscosity, $k_B$ the Boltzmann constant, and $T$ the
temperature.~\cite{anderson:89,colberg14}

The Reynolds number is assumed to be small so that viscous forces dominate inertial forces and the fluid flow field outside of the boundary layer is found by solving the Stokes equation with the incompressibility condition,
\begin{equation}
  \nabla p = \bar{\mu} \nabla^2 \bm{v},
  \hspace{30pt} \nabla \cdot \bm{v} = 0,
\end{equation}
subject to the boundary conditions in the laboratory frame of reference,
\begin{equation}
  \bm{v}_{\eta=\eta_1} = 0, \quad   \bm{v}_{\eta=\eta_2} = (\bm{V} + \bm{v}_s)_{\eta=\eta_2},
  \label{bd_fluid}
\end{equation}
where $p$ is the pressure, $\bm{v}$ the fluid velocity field,
and $\bm{V}$ the velocity of the noncatalytic sphere.

Introducing the stream function $\psi$, which is related to the flow velocity by
$\bm{v}=\hat{\bm{\phi}}/\rho \times \nabla \psi$, where
$\rho=\xi \sin \theta/(\cosh \eta -\mu)$,
one may replace the Stokes equation with the incompressibility condition in
terms of stream functions by~\cite{happel73,jeff26}
\begin{align}
E^4 (\psi) = 0,
\label{eq:stream_eq}
\end{align}
where $E^4=E^2(E^2)$ and
$E^2 = (\cosh\eta-\mu)/\xi^2
[\partial / \partial \eta \{ (\cosh\eta-\mu)\partial / \partial \eta \}
+(1-\mu^2) \partial / \partial \mu
\{ (\cosh\eta-\mu) \partial / \partial \mu \} ] $.
This equation has an exact solution given by~\cite{jeff26}
\begin{align}
  \psi = (\cosh\eta-\mu)^{-\frac{3}{2}}\sum_{n=1}^{\infty}W_n(\eta)V_n(\mu),
  \label{eq:stream}
\end{align}
where
$W_n(\eta) = a_n\cosh(n-\tfrac{1}{2})\eta
+ b_n\sinh(n-\tfrac{1}{2})\eta
+c_n\cosh(n+\tfrac{3}{2})\eta
+ d_n\sinh(n+\tfrac{3}{2})\eta$
and $V_n(\mu) = P_{n-1}(\mu) - P_{n+1}(\mu)$. The unknown coefficients $a_n$, $b_n$, $c_n$, and $d_n$ in Eq.~(\ref{eq:stream})
are determined by boundary conditions at the outer edges of the boundary
layers around the $S_1$ and $S_2$ spheres, i.e. Eq.~(\ref{bd_fluid}). In
the laboratory frame where the motor moves with velocity $V$, these boundary
conditions are given in terms of the stream function by
\begin{eqnarray}
&&\psi \vert_{\eta=\eta_1} = 0, \quad (\psi + \tfrac{1}{2}\rho^2 V)\vert_{\eta=\eta_2} = 0, \nonumber \\
&&\frac{\partial \psi}{ \partial \eta} \bigg \vert_{\eta=\eta_1} = 0, \quad
\frac{\partial}{ \partial \eta} (\psi + \tfrac{1}{2}\rho^2 V) \bigg \vert_{\eta=\eta_2}
= \kappa\rho \frac{\partial c_B}{ \partial \theta } \bigg \vert_{\eta=\eta_2}.
\label{bd_psi}
\end{eqnarray}

By writing $\chi = \sum_{n=1}^{\infty}W_n(\eta)V_n(\mu)$ in Eq.~(\ref{eq:stream}),
we can replace the boundary conditions, Eq.~(\ref{bd_psi}) in terms of $\chi$
by
\begin{align}
  \chi \vert_{\eta=\eta_1}  &= 0,  \hspace{20pt}
  \frac{\partial\chi}{\partial \eta}\bigg\vert_{\eta=\eta_1} = 0,  \nonumber \\
  \chi \vert_{\eta=\eta_2}  &= -\frac{\xi^2 V (1-\mu^2)}{2(\cosh\eta-\mu)^{1/2}}\bigg\vert_{\eta=\eta_2}, \nonumber \\
%  \frac{\partial\chi}{\partial \eta}\bigg\vert_{\eta=\eta_1} &= 0, \nonumber
%  \\
  \frac{\partial\chi}{\partial \eta}\bigg\vert_{\eta=\eta_2}
  &= \frac{\xi^2 V (1-\mu^2)\sinh\eta}{4(\cosh\eta-\mu)^{3/2}}\bigg\vert_{\eta=\eta_2} %\nonumber\\
  +\xi\kappa \sum_{n=0}^{\infty} \Big[A_n e^{(n+\frac{1}{2})\eta} \nonumber\\
  &\hspace{20pt}+B_n e^{-(n+\frac{1}{2})\eta}\Big]
  \bigg[-\frac{(1-\mu^2)P_n}{2} \nonumber \\
  &\hspace{20pt}+(\cosh\eta-\mu)(1-\mu^2)\frac{d P_n}{d \mu}\bigg]
  \bigg\vert_{\eta=\eta_2}.
  \label{eq:xi}
\end{align}
Here, $1/\sqrt{\cosh\eta-\mu}$ can be expressed in a series of Legendre
function $P_n$,
$(1-\mu^2)P_n$ and $\mu V_n$ are rewritten by Gegenbauer functions $V_{n-1}$ and $V_{n+1}$,
and $(1-\mu^2)dP_n/d\mu$ is rewritten by $V_n$.~\cite{jeff26,grad07,reigh:15dimer}
Then, we may expand the right sides of
Eq.~(\ref{eq:xi}) for $\eta=\eta_2$ in a series of $V_n$ as
\begin{align}
  &\chi \vert_{\eta=\eta_2}  = -\frac{\xi^2 V}{\sqrt{2}}
  \sum_{n=1}^{\infty} \frac{n(n+1)}{2n+1}\bigg[ \frac{e^{(n-1/2)\eta_2}}{2n-1}
  - \frac{e^{(n+3/2)\eta_2}}{2n+3} \bigg] V_n, \nonumber \\
  &\frac{\partial\chi}{\partial \eta}\bigg\vert_{\eta=\eta_2}
  = -\frac{\xi^2 V}{2\sqrt{2}}\sum_{n=1}^{\infty}\frac{n(n+1)}{2n+1}
  [e^{(n-1/2)\eta_2}-e^{(n+3/2)\eta_2}]V_n \nonumber\\
  &\hspace{50pt}+ \xi\kappa\sum_{n=1}^{\infty}\Phi_n V_n.
  \label{eq:xi1}
\end{align}

Since both sides of Eqs.~(\ref{eq:xi1}) are expanded in a series of
Gegenbauer function $V_n$,
we can determine the unknown coefficients of $W_n(\eta)$ in Eq.~(\ref{eq:stream})
from the following equations:
\begin{align}
  &a_n \cosh(n-\tfrac{1}{2})\eta_1 + b_n \sinh(n-\tfrac{1}{2})\eta_1 \nonumber\\
  &\quad +c_n \cosh(n+\tfrac{3}{2})\eta_1 + d_n \sinh(n+\tfrac{3}{2})\eta_1 \nonumber\\
  &\qquad = 0,  \nonumber\\
%%%%%%%%%%%%%%%%%%%%%
%-\gamma_l\{(2l+3)e^{-(l-\frac{1}{2})\eta_1}-(2l-1)e^{-(l+\frac{3}{2})\eta_1} \}, \nonumber\\
  &a_n \cosh(n-\tfrac{1}{2})\eta_2 + b_n \sinh(n-\tfrac{1}{2})\eta_2 \nonumber\\
  &\quad +c_n \cosh(n+\tfrac{3}{2})\eta_2 + d_n \sinh(n+\tfrac{3}{2})\eta_2 \nonumber\\
  &\qquad =
  -\gamma_n\{(2n+3)e^{(n-\frac{1}{2})\eta_2}-(2n-1)e^{(n+\frac{3}{2})\eta_2}
  \}, \nonumber\\
%%%%%%%%%%%%%%%%%%%%%
  &(2n-1) \{a_n \sinh(n-\tfrac{1}{2})\eta_1 + b_n \cosh(n-\tfrac{1}{2})\eta_1 \}\nonumber\\
  &\quad +(2n+3)\{c_n \sinh(n+\tfrac{3}{2})\eta_1 + d_n \cosh(n+\tfrac{3}{2})\eta_1 \} \nonumber\\
  &\qquad = 0, \nonumber\\
%  &(2l-1)(2l+3)\gamma_l\{e^{-(l-\frac{1}{2})\eta_1}-e^{-(l+\frac{3}{2})\eta_1}
%\} \nonumber\\
%\end{align}
%\begin{align}
%%%%%%%%%%%%%%%%%%%%%%%%%%%%%%%%%%%
  &(2n-1) \{a_n \sinh(n-\tfrac{1}{2})\eta_2 + b_n \cosh(n-\tfrac{1}{2})\eta_2 \}\nonumber\\
  &\quad +(2n+3)\{c_n \sinh(n+\tfrac{3}{2})\eta_2 + d_n \cosh(n+\tfrac{3}{2})\eta_2 \} \nonumber\\
  &\qquad = -(2n-1)(2n+3)\gamma_n\{e^{(n-\frac{1}{2})\eta_2}-e^{(n+\frac{3}{2})\eta_2} \}\nonumber\\
  &\hspace{30pt} +2\xi\kappa\Phi_n,
\end{align}
where $\gamma_n=f_n V$ and $f_n$ is given in Table~\ref{tab1} in the Appendix.
The solution of the above equations for the unknown coefficients $a_n$, $b_n$,
$c_n$, $d_n$ is expressed by
\begin{align}
  \Delta_n \bm{X} = \gamma_n \bm{Y}^{(e)} -\frac{1}{2}\xi \kappa \Phi_n \bm{Z},
  \label{eq:coeff1}
\end{align}
where $\bm{X}=\{a_n,b_n,c_n,d_n\}$,
$\bm{Y}^{(e)}=\{Y_n^{(2)}, Y_n^{(4)}, Y_n^{(6)}, Y_n^{(8)}\}$, and
$\bm{Z}=\{z_n^{(1)},z_n^{(2)},z_n^{(3)},z_n^{(4)} \}$.
The elements of the vectors are given in Table~\ref{tab1}.
The solution for two inactive spheres can be obtained easily by taking
$\kappa=0$, which gives $\bm{X}=\gamma_n \bm{Y}^{(e)}/\Delta_n$.
In this case, one colloidal sphere ($S_2$) with constant velocity $V$ moves to
the other sphere ($S_1$) fixed in space.

The forces ($F_1, F_2$) on the individual spheres ($S_1, S_2$) are given by
integrating the stress on the surface of the boundary layer,
$F_i = \int_{S_i} \bm{\Pi}_{i,z} \cdot \hat{\bm{n}} dS_i$ ($i=1,2$),
where $\bm{\Pi}_{i,z} = \hat{\bm{z}} \cdot \bm{\Pi}_i$
and $\bm{\Pi}$ is the stress tensor.
The system is symmetric around the azimuthal angle $\phi$ and only the
force in the $z$-direction needs to be considered.
The analytic expressions for the force exerted on the spheres by the fluid
are given in Stimson and Jeffery~\cite{jeff26} as
\begin{align}
  F_1 &= \frac{2\sqrt{2}\pi\bar{\mu}}{\xi}\sum_{n=1}^{\infty}(2n+1)(a_n+b_n+c_n+d_n), \nonumber\\
  F_2 &=
  \frac{2\sqrt{2}\pi\bar{\mu}}{\xi}\sum_{n=1}^{\infty}(2n+1)(a_n-b_n+c_n-d_n).
  \label{eq:force_fix}
\end{align}

The velocity can be found from these force expressions.
Since no external force is applied to the $S_2$ sphere, although
the $S_1$ sphere is fixed in space by an external force,
the total force on the $S_2$ sphere  at the outer edge of the boundary layer
is zero, $F_2=0$.
Noting that $\gamma_n=f_n V$,
one can find the following expression for velocity of the noncatalytic sphere,
\begin{align}
  V = \kappa \frac{\xi}{2}\frac{\displaystyle\sum_{n=1}^{\infty}(2n+1)
    \Phi_n \Xi_n^{(-)} /\Delta_n}
  {\displaystyle\sum_{n=1}^{\infty}(2n+1)f_n\Gamma_n^{(+)} /\Delta_n}.
%    \equiv \kappa {\cal C}^{{\rm F}}(L).
  \label{mvel_fix}
\end{align}
Also, the force $F_1$ exerted on the fixed catalytic sphere by the fluid found
here is used for the plots in Fig.~\ref{fig:force}.

\subsubsection{Freely moving catalytic sphere}
We now suppose that both spheres are free to move and construct the solutions for this force-free case.
Letting the velocities of the $S_1$ and $S_2$ spheres be $\bm{V}^{(1)}$ and
$\bm{V}^{(2)}$, respectively, one may replace the boundary conditions in Eq.~(\ref{bd_fluid}) by
\begin{equation}
  \bm{v}_{\eta=\eta_1} = (\bm{V}^{(1)})_{\eta=\eta_1}, \quad
  \bm{v}_{\eta=\eta_2} = (\bm{V}^{(2)} + \bm{v}_s)_{\eta=\eta_2}.
\end{equation}
Then the boundary conditions for the stream function are
\begin{eqnarray}
  (\psi + \tfrac{1}{2}\rho^2 V^{(i)})\vert_{\eta=\eta_i} &=& 0, \nonumber \\
  \frac{\partial}{ \partial \eta} (\psi + \tfrac{1}{2}\rho^2 V^{(i)}) \bigg \vert_{\eta=\eta_i}
  &=& \kappa\rho \frac{\partial c_B}{ \partial \theta } \bigg
  \vert_{\eta=\eta_i}\Theta_i,
  \label{bd_mov}
\end{eqnarray}
where $\Theta_1=0$, $\Theta_2=1$, and $i=1,2$.

In this case, the boundary conditions for streamlines in Eq.~(\ref{bd_mov})
are rewritten in terms of $\chi = \sum_{n=1}^{\infty}W_n(\eta)V_n(\mu)$ by
\begin{align}
  &\chi \vert_{\eta=\eta_i}  = -\frac{\xi^2 V^{(i)}
    (1-\mu^2)}{2(\cosh\eta-\mu)^{1/2}}\bigg\vert_{\eta=\eta_i}, \nonumber \\
  &\frac{\partial\chi}{\partial \eta}\bigg\vert_{\eta=\eta_i}
  = \frac{\xi^2 V^{(i)}
    (1-\mu^2)\sinh\eta}{4(\cosh\eta-\mu)^{3/2}}\bigg\vert_{\eta=\eta_i} \nonumber\\
  &\hspace{40pt}+\xi\kappa \sum_{n=0}^{\infty} \Big[A_n e^{(n+\frac{1}{2})\eta} 
  +B_n e^{-(n+\frac{1}{2})\eta}\Big] \times \nonumber\\
  &\hspace{50pt}\bigg[-\frac{(1-\mu^2)P_n}{2}  \nonumber\\
  &\hspace{50pt}+(\cosh\eta-\mu)(1-\mu^2)\frac{d P_n}{d \mu}\bigg]
  \bigg\vert_{\eta=\eta_i} \Theta_i,
  \label{eq:xi_two}
\end{align}
where $\Theta_1=0$, $\Theta_2=1$, and $i=1,2$.

As discussed previously,
%Using Eqs.~(\ref{relation}) and~(\ref{expan1}),
we may expand the right sides of Eq.~(\ref{eq:xi_two}) in a series of
Gegenbauer function $V_n$ as
\begin{align}
  &\chi \vert_{\eta=\eta_i}  = -\frac{\xi^2 V^{(i)}}{\sqrt{2}}
  \sum_{n=1}^{\infty} \frac{n(n+1)}{2n+1} \nonumber\\
  &\hspace{80pt} \times \bigg[ \frac{e^{\mp(n-1/2)\eta_i}}{2n-1}
  - \frac{e^{\mp(n+3/2)\eta_i}}{2n+3} \bigg] V_n, \nonumber \\
  &\frac{\partial\chi}{\partial \eta}\bigg\vert_{\eta=\eta_i}
  = \pm \frac{\xi^2 V^{(i)}}{2\sqrt{2}}\sum_{n=1}^{\infty}\frac{n(n+1)}{2n+1} \nonumber\\
  &\hspace{80pt} \times [e^{\mp(n-1/2)\eta_i}-e^{\mp(n+3/2)\eta_i}]V_n \nonumber\\
  &\hspace{50pt}+ \bigg(\xi\kappa\sum_{n=1}^{\infty}\Phi_n V_n \bigg) \Theta_i,
  \label{eq:xi2}
\end{align}
where the upper and lower signs are taken for $i=1$ and $2$, respectively.

Since both sides of Eqs.~(\ref{eq:xi2}) are expanded in a series of $V_n$,
we can determine the unknown coefficients of $W_n(\eta)$ in Eq.~(\ref{eq:stream})
from the following equations:
\begin{align}
  &a_n \cosh(n-\tfrac{1}{2})\eta_i + b_n \sinh(n-\tfrac{1}{2})\eta_i \nonumber\\
  &\quad +c_n \cosh(n+\tfrac{3}{2})\eta_i + d_n \sinh(n+\tfrac{3}{2})\eta_i \nonumber\\
  &\qquad =
  -\gamma_n^{(i)}\{(2n+3)e^{\mp (n-\frac{1}{2})\eta_i}-(2n-1)e^{\mp (n+\frac{3}{2})\eta_i}
  \}, \nonumber\\
%%%%%%%%%%%%%%%%%%%%%
%%%%%%%%%%%%%%%%%%%%%%%%%%%%%%%%%%%
  &(2n-1) \{a_n \sinh(n-\tfrac{1}{2})\eta_i + b_n \cosh(n-\tfrac{1}{2})\eta_i \}\nonumber\\
  &\quad +(2n+3)\{c_n \sinh(n+\tfrac{3}{2})\eta_i + d_n \cosh(n+\tfrac{3}{2})\eta_i \} \nonumber\\
  &\qquad = \pm (2n-1)(2n+3)\gamma_n^{(i)}\{e^{\mp
    (n-\frac{1}{2})\eta_i}-e^{\mp (n+\frac{3}{2})\eta_i} \}\nonumber\\
  &\qquad +2\xi\kappa\Phi_n \Theta_i,
\end{align}
where $\gamma_n^{(i)}=f_n V^{(i)}$ and the upper and lower signs correspond to $i=1$ and $2$, respectively.

The solution of the above equations for the unknown coefficients $a_n$, $b_n$, $c_n$, $d_n$ is given by
\begin{align}
  \Delta_n \bm{X} = \gamma_n^{(1)} \bm{Y}^{(o)} + \gamma_n^{(2)} \bm{Y}^{(e)}
  -\frac{1}{2}\xi \kappa \Phi_n \bm{Z},
  \label{eq:coeff2}
\end{align}
where $\bm{X}=\{a_n,b_n,c_n,d_n\}$,
$\bm{Y}^{(o)}=\{Y_n^{(1)}, Y_n^{(3)}, Y_n^{(5)}, Y_n^{(7)} \}$,
$\bm{Y}^{(e)}=\{Y_n^{(2)}, Y_n^{(4)}, Y_n^{(6)}, Y_n^{(8)}\}$, and
$\bm{Z}=\{z_n^{(1)},z_n^{(2)},z_n^{(3)},z_n^{(4)} \}$.
The elements of the vectors are given in Table~\ref{tab1}.
Applying the force-free conditions on both the spheres, $F_1=F_2=0$ in
Eq.~\ref{eq:force_fix}, one can find the solution for the velocities of the $S_1$ and $S_2$ spheres as
\begin{align}
  V^{(1)} &= -\frac{\mathcal{A}^{(0)} \mathcal{B}^{(-)} - \mathcal{A}^{(+)}\mathcal{B}^{(+)}}{\mathcal{A}^{(+)} \mathcal{A}^{(-)}
    - (\mathcal{A}^{(0)})^2},
  \nonumber\\
  V^{(2)} &= \frac{\mathcal{A}^{(-)} \mathcal{B}^{(-)} - \mathcal{A}^{(0)}\mathcal{B}^{(+)}}{\mathcal{A}^{(+)} \mathcal{A}^{(-)}
    - (\mathcal{A}^{(0)})^2},
  \label{eq:mvel_mov}
\end{align}
where
\begin{align}
  &\mathcal{A}^{(\pm,0)} =
  \displaystyle\sum_{n=1}^{\infty}(2n+1)f_n\Gamma_n^{(\pm,0)}/\Delta_n,  \nonumber\\
  &\mathcal{B}^{(\pm)} =\kappa \frac{1}{2}\xi
  \displaystyle\sum_{n=1}^{\infty}(2n+1)\Phi_n\Xi_n^{(\pm)}/\Delta_n.
\end{align}

The solutions for two inactive spheres moving with constant velocities
$V^{(1)}$ and $V^{(2)}$ along the axisymmetric direction can be obtained easily by setting
$\kappa=0$, which gives $\bm{X}= (\gamma_n^{(1)}\bm{Y}^{(o)}+\gamma_n^{(2)}\bm{Y}^{(e)})/\Delta_n$.
Also, the solutions for the sphere-dimer can be obtained by setting
$V^{(1)}=V^{(2)}=V$, which gives
$\Delta_n \bm{X} = \gamma_n \bm{Y} -\frac{1}{2}\xi \kappa \Phi_n \bm{Z}$,
where $\bm{Y}=\bm{Y}^{(o)}+\bm{Y}^{(e)}$ and
$\gamma_n=\gamma_n^{(1)}=\gamma_n^{(2)}$.
This expression is consistent with the formula given earlier.~\cite{reigh:15dimer,footnote:typo}

\section{\label{sec:sim}Microscopic dynamics}

The analytical results for continuum theory are exact given the formulation of the problem on
which they are based. In particular, they rest on the deterministic   continuum description of the fluid and
solute concentration as described by the Stokes and diffusion equations, supplemented with boundary conditions on the fluid velocity and concentration fields. The former boundary condition accounts for the fluid dynamics and the
latter boundary condition describes chemical reactions on the sphere. The fluid viscosity, diffusion constant and reaction rates of chemical species are specified as input parameters to solve the equations. The Reynolds and P\'{e}clet numbers are asummed to be small.~\cite{anderson:89,golestanian:05,pope11,michelin:15,reigh:15dimer}
This is an appropriate description for a large macroscopic particle. However, in many experiments, the active particles have micrometer or nanometer dimensions and for such systems thermal fluctuations should be
taken into account.~\cite{palacci:13,buttinoni:13,fischer:14,singh:17,schmidt:18,Yu:18}
In addition, as one moves to small nanometer~\cite{fischer:14} or even Angstrom~\cite{colberg14} scales the assumptions of continuum dynamics may no longer apply.

The coarse-grain particle-based simulations do not make such assumptions. The input parameters are the intermolecular potentials and multiparticle collision parameters for the solvent.~\cite{footnote:mpc1}
The resulting dynamics then yields all other properties such as the transport coefficients of the system, and other dimensionless numbers that characterize the system. One can show that on long distance and times scales the continuum hydrodynamic and diffusion equations are recovered~\cite{malevanets:99}, but the dynamics is not restricted to this
limit. Consequently, it is of interest to examine the extent to which the continuum model can capture the active dynamics of these small particles.~\cite{reigh:15dimer,reigh:16janus}

The coarse-grain microscopic dynamics we employ combines molecular dynamics (MD) with multiparticle collision (MPC) dynamics.~\cite{malevanets:99,malevanets:00} More specifically, the fluid is composed of $N_s$ point particles of mass $m$ with positions
${\bm{r}}_i$ and velocities $\bm{v}_i$, where $i=1,\ldots,N_s$. There are no explicit intermolecular potentials among these fluid particles and their interactions are accounted for by multiparticle collisions. The dynamics consists of two alternating steps: streaming and collision. In the streaming steps of duration $h$, all particles in the system move by Newton's equations of motion with forces determined by the sphere-sphere and sphere-solvent intermolecular potentials. At each collision time the solvent particles are sorted into cubic cells of side length $a$, which is larger than the mean free path, and their relative velocities are rotated around a randomly oriented axis by a fixed angle $\alpha$ with respect to the center-of-mass
velocities of each cell. The velocity of particle $i$ after collision is given by
$\bm{v}_i(t+h)=\bm{v}_{cm}(t)  +\mathcal{R}(\alpha)(\bm{v}_i(t)-{\bm
  v}_{cm}(t))$, where $\mathcal{R}(\alpha)$ is the rotation matrix, ${\bm
  v}_{cm}=\Sigma^{N_c}_{j=1} {\bm v}_j/N_c$ is the center-of-mass velocity of the particles in the cell to which the particle $i$ belongs, and
$N_c$ is the number of particles in that cell. A random shift of the collision
lattice is applied at every collision step to ensure Galilean
invariance.~\cite{ihle:01}
The dynamics locally conserves mass, momentum and energy.~\cite{footnote:mpc1}

The spheres interact with the fluid particles through repulsive Lennard-Jones (LJ) potentials, $
U= 4 \epsilon [ \left( \sigma/r \right)^{12}-  \left( \sigma/r \right)^6 ]+\epsilon$ for $r<2^{1/6}\sigma$ and $U=0$ for $r \geq 2^{1/6}\sigma$ with energy $\epsilon$ and distance $\sigma$ parameters. In addition, repulsive LJ potentials are employed to take into account excluded volume interactions between the two spheres with $\sigma_s$ denoting the value of $\sigma$ in this case. In order to make only the noncatalytic sphere hydrodynamically active, we
choose the interaction energies of the $A$ and $B$ molecules with the $S_1$ catalytic
sphere to be the same ($\epsilon_A=\epsilon_B=\epsilon$) and those with the
$S_2$ noncatalytic sphere to be different ($\epsilon_B <
\epsilon_A=\epsilon$). Setting $\epsilon_B < \epsilon_A$, so that the $A$
particles are more strongly repelled from the $S_2$ sphere than the $B$
particles, causes it to move towards the $S_1$ sphere; hence $B$ plays the
role of chemoattractant. An irreversible chemical reaction $A \rightarrow B$
takes place on the $S_1$ sphere with intrinsic reaction rate $k_0$ whenever
$A$ encounters $S_1$. Collisions of $A$ or $B$ particles with the $S_2$ sphere
do not lead to reaction. To maintain the system in a steady state, the $B$
particles are converted to $A$ at a distance $d_p=L_b/2$ far from the
spheres.

All quantities are reported in dimensionless units where length, energy, mass and time are measured in units of the MPC cell length $a=\sigma/2$, $\epsilon$, the solvent mass $m$, and $a\sqrt{m/\epsilon}$, respectively.
The cubic simulation box with linear dimension $L_b=50$ and
periodic boundary conditions in all dimensions is divided into $L_b^3 =
50^3$ cubic cells. Multiparticle collisions are carried out in each cell by performing
velocity rotations by an angle $\alpha=120^\circ$ about a randomly chosen axis every collision time $h = 0.1$.
The average solvent number density is $c_0=10$ and
the temperature is $k_BT=1$. The MD time step is $\Delta t = 0.01$. The energy
parameters for the $S_2$ sphere-fluid repulsive LJ potentials are $\epsilon_A=1.0$ and $\epsilon_B=0.1$ for $A$ and $B$, respectively, while
$\epsilon_A=\epsilon_B=1.0$ for the $S_1$ sphere.
The size parameters are $\sigma=2$ and $\sigma_s=4$ to give effective sphere radii of $R_1=R_2=2^{1/6}\sigma$.
The sphere mass is taken to be $M=4\pi \sigma^3 c_0/3$ corresponding to neutral buoyancy.
The intrinsic reaction rate constant for the $A + S_1 \rightarrow B + S_1$ reaction can be estimated from simple collision
theory so that $\bar{k}_0\sim \sqrt{k_BT/2\pi m} \sim 0.4$.
The transport properties of the fluid depend on $h$, $\alpha$, and $N_c$.
The fluid viscosity is $\bar{\mu} = m N_c \nu = 7.9$, where $\nu$ \
is the kinematic viscosity, and the common $A$ and $B$ diffusion constant is $D=0.0611$. The Schmidt number is $S_c=\nu/D=13>1$, which ensures that momentum transport dominates over mass transport,
the Reynolds number $R_e= c_0Va/\bar{\mu} < 0.1$, implying that viscosity is
dominant over inertia, and the P\'{e}clet number $Pe = Va/D <1$, diffusion being dominant over fluid advection.

The parameter values given above are used as input
to obtain the analytic solutions in the continuum theory.
For example, the factor $\kappa$ in Eq.~(\ref{eq:kappa}) is obtained
from the repulsive cut-off LJ potentials with the energy parameters
$\epsilon_A$ and $\epsilon_B$ given in simulations, along with the viscosity from the microscopic model.
Using the analytical continuum solutions and simulations of the microscopic equations of motion, we can discuss the physics underlying dynamics of these two-sphere systems. Since the phenomena depend on whether the catalytic sphere is fixed or free to move, we discuss these two cases separately.

\section{Dynamics with a fixed catalytic sphere}\label{sec:fix}

The process by which a noncatalytic sphere responds to the chemical gradient produced by a fixed catalytic sphere and is captured by it has been studied earlier.~\cite{snigdha:14,tao:16} Here we reexamine this process by making use of analytical solutions and extensive simulations of the microscopic model. The dynamical processes that enter this seemingly simple process involve effects that govern the velocity of the noncatalytic sphere and lead to its eventual capture. At large radial distances between the spheres the concentration of product $B$ in the vicinity of $S_2$ is low and so is its velocity. As the distance decreases the concentration of $B$ increases leading to an increased velocity but as the spheres approach closely more complex interactions lead to the capture event. We are able to probe the details of the mechanism responsible for the capture process through an analysis of the concentration and fluid flow fields that accompany the dynamics.

\begin{figure}[t]%[htbp]
  \centering
  \includegraphics[scale=0.55,angle=0]{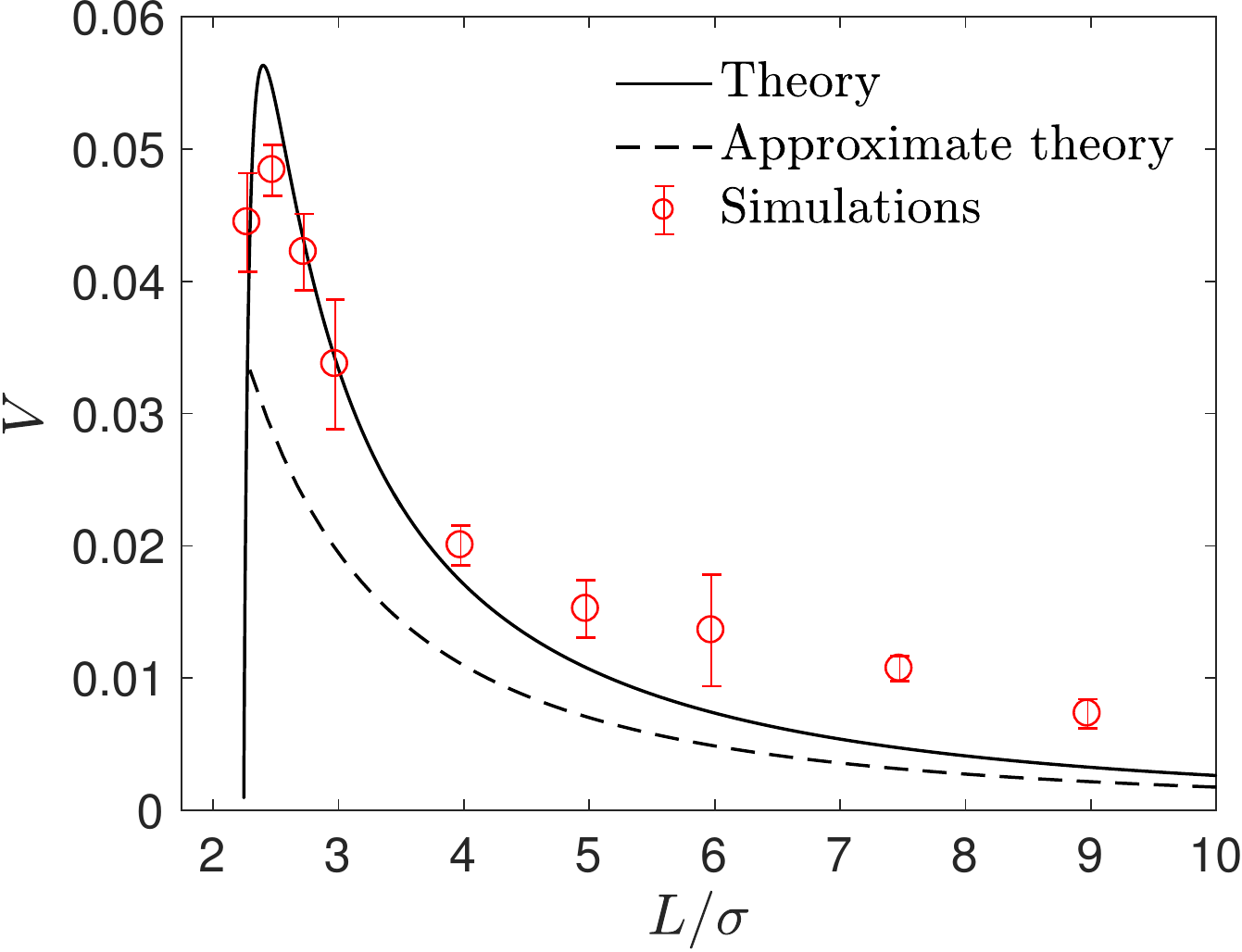}
  \caption{The velocity $V$ of the noncatalytic $S_2$ sphere as a function of
    the separation $L$ between the $S_2$ and $S_1$ catalytic spheres.
    The black solid line is the exact solution calculated from continuum
    hydrodynamic theory in Eq.~(\ref{mvel_fix}) and the black dashed line is the approximate velocity $V_a$
    from Eq.~(\ref{approx}) that is valid for large $L$.  The red circles with
    error bars are the results of microscopic simulations. Averages were
    obtained from 80 realizations of the dynamics.  }
  \label{fig:vel_r}
\end{figure}

The velocity of the $S_2$ sphere, $V$, is plotted in Fig.~\ref{fig:vel_r} as a
function of the distance $L$ separating the centers of the two spheres. The
figure shows the expected increase in velocity as the $S_2$ sphere approaches
the $S_1$ sphere until, at a short distance, it begins to decrease as the
capture event takes place. The figure compares the simulation results with the
exact analytical continuum theory result in Eq.~(\ref{mvel_fix}). The results
are also compared with an approximate theory
where the two spheres are assumed to be separated by a large distance.
In this case, the concentration field may be approximated by calculating it
in the absence of the $S_2$ sphere~\cite{snigdha:14,tao:16} as follows. Taking the origin of a spherical polar coordinate ($r_1, \theta_1,\phi_1$)
at the center of the $S_1$ sphere in Fig.~\ref{fig:coord}, the $B$ species concentration field may be obtained from the solution of
the diffusion equation~(\ref{diff}) subject to the radiation boundary condition in Eq.~(\ref{bound}) as
\begin{eqnarray}
%  c_A(r_1) = c_0 \Big\{ 1 - \frac{k_0}{k_0+k_D}\frac{R_1}{r_1} \Big\}.
  c_B(r_1) =  \frac{c_0k_0}{(k_0+k_D)}\frac{R_1}{r_1},
  \label{conc_app}
\end{eqnarray}
where $k_D=4\pi R_1 D$ is the Smoluchowski rate coefficient.
This far-field concentration field can be also obtained from the
  approximation of the exact solution of the two spheres in large distance,
  $c_B=-\sqrt{2}\xi\sum_{n=0}^{\infty}(A_n+B_n)/r +
  \mathcal{O}(1/r^2)$,~\cite{reigh:15dimer}
  where a new spherical polar coordinate ($r$, $\vartheta$, $\phi$) in 
  Fig.~\ref{fig:coord} is chosen sharing the origin,  
  by taking the limit of $\eta_2 \rightarrow -\infty$ ($R_2 \rightarrow 0$) and
$L \rightarrow \infty$ and noting that the $n=0$ term is sufficient.

The approximation to the propulsion velocity of the $S_2$ sphere may be then found
by averaging the slip velocity like Eq.~(\ref{slip_vel}) at the edge of the
boundary layer of the $S_2$ sphere~\cite{anderson:89,stone:96,colberg14} in a coordinate system
($r_2, \theta_2, \phi_2$) where the origin is at the center of the $S_2$
sphere.
The result is
\begin{eqnarray}
  V_a = -\frac{1}{4\pi R_2^2} \int_{S_2} \bm{v}_s \cdot \hat{\bm{z}} dS_2.
  \label{approx}
\end{eqnarray}
Here, $\hat{\bm{z}}$ is a unit vector along the line of centers of the two spheres and defines the $z$-axis of the spherical polar coordinate system. Using the relation $r_1^2 = r_2^2 + L^2 -2r_2L\cos\theta_2$,
one obtains $c_B(r_2)$ from Eq.~\ref{conc_app}
and hence an approximate expression for the sphere velocity for distances $L \gg R_2$ given by
\begin{align}
  V_a = \frac{2 \kappa c_0k_0R_1}{3(k_0+k_D)L^2}.
  \label{appr_vel}
\end{align}
As expected, the approximate and exact theories agree for large sphere separations where both have a $L^{-2}$ power law behavior, but significant deviations are seen a short distances. The discrepancies between the microscopic simulations and exact continuum theory may be due to the use of soft potential functions and features of microscopic dynamics taking place in the boundary layer which are not captured by the simple boundary conditions in the continuum model, which likely manifest themselves more strongly at large separations where the product concentrations and gradients are small.

\begin{figure}[t]
  \centering
  \includegraphics[scale=0.8,angle=0]{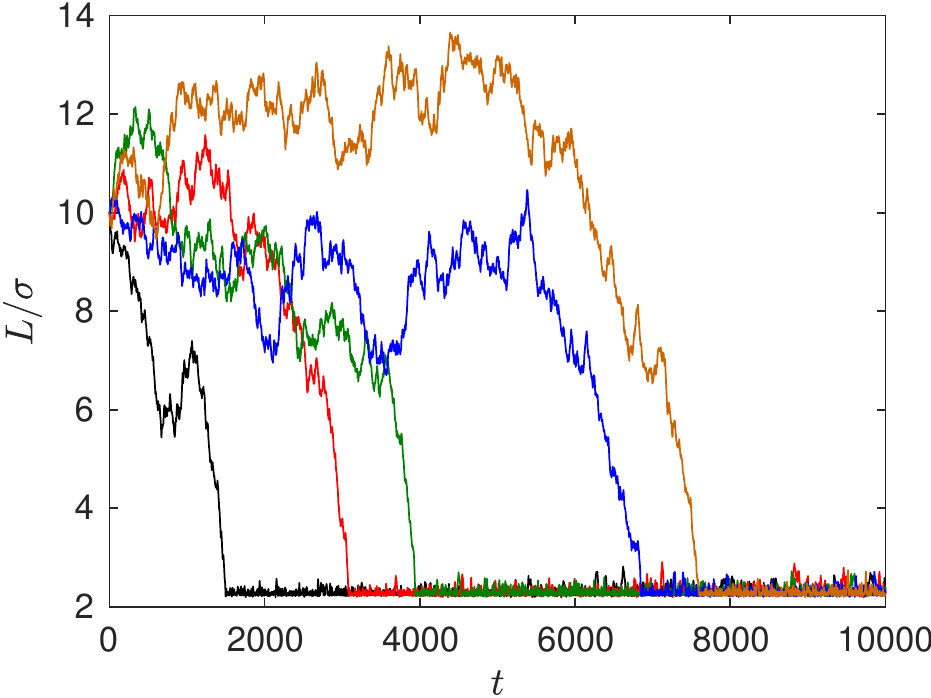}
  \caption{Plot of the distance between the fixed catalytic and moving noncatalytic spheres as a function of time.
      Five realizations of the dynamics are shown, each with an initial separation $L/\sigma = 10$. Contact occurs at approximately $L/\sigma \sim 2.3$. The time where the distance achieves its minimum value is the capture time (see Fig.~\ref{fig:captm_fix}).}
    \label{fig:dist_tm}
\end{figure}
\begin{figure}[t]
  \centering
  \includegraphics[scale=0.8,angle=0]{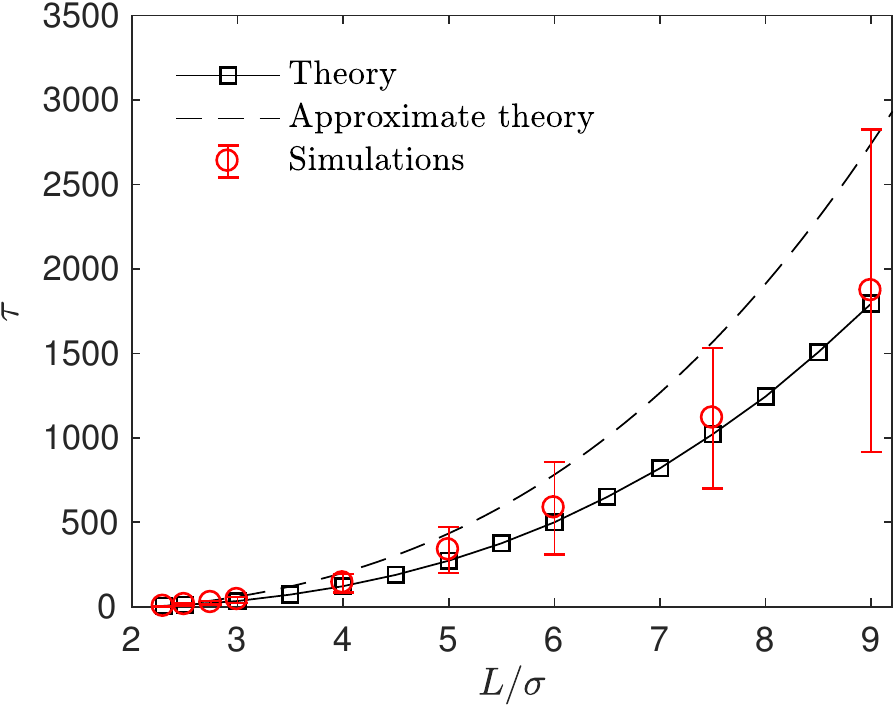}
  \caption{Capture time $\tau$ as a function of the initial separation $L$ between the spheres.
    The black solid line with squares is the exact continuum solution and the dashed line is the approximate result.
    The red circles denote the simulation results obtained from averages over
    80 realizations.
   }
  \label{fig:captm_fix}
\end{figure}
\begin{figure}[t!]
  \centering
  \includegraphics[scale=0.47,angle=0]{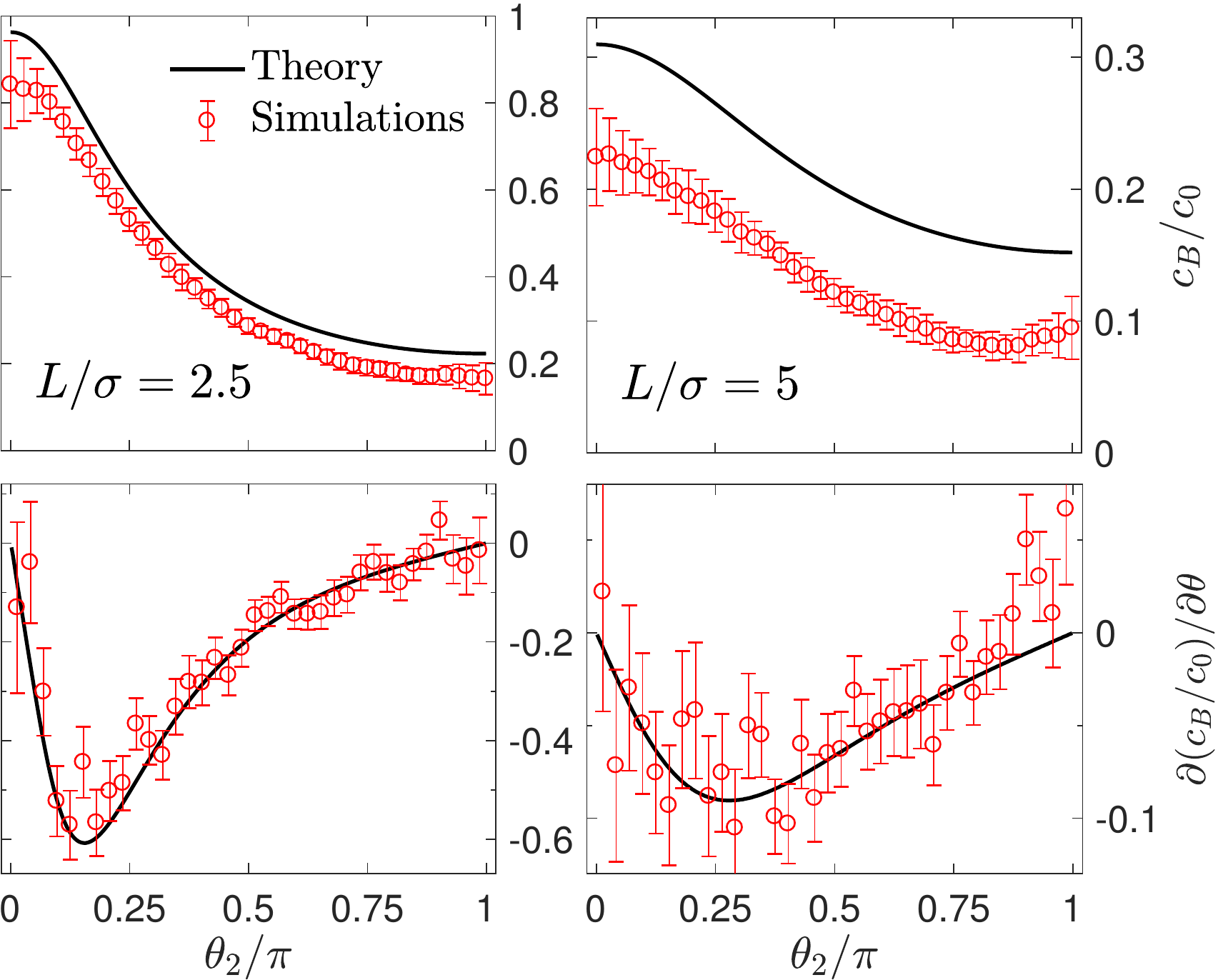}
  \caption{
    Normalized concentration fields $c_B/c_0$ and the tangential gradients
      $\partial{(c_B/c_0)}/\partial \theta_2$ on the surface of the noncatalytic
      $S_2$ sphere ($R_2/\sigma=2^{1/6}$) for $L/\sigma=2.5$ (left column) and
      $L/\sigma=5$ (right column), respectively.
      The angle $\theta_2$ is the polar angle in spherical polar
      coordinates where the origin is at the center of the $S_2$ sphere.
      At $\theta_2=0$, the $+\theta_2$ direction is the $+z$ direction in Fig.~\ref{fig:coord}.}
  \label{fig:conc_fd}
\end{figure}

In the microscopic simulations the colloidal particles undergo Brownian motion as a result of thermal fluctuations, as well as directed motion due to diffusiophoresis. Figure~\ref{fig:dist_tm} shows some examples of noncatalytic sphere trajectories.
At large distances ($L/\sigma > 8$) the noncatalytic sphere exhibits small thermal fluctuations in its displacement which are less than its radius, as well as larger random displacements. When $L/\sigma < 6$, diffusiophoretic interactions are stronger and the deterministic component of the motion dominates. Thus, fluctuations lead to a dispersion of capture times seen in Fig.~\ref{fig:dist_tm}, and only the average in Fig.~\ref{fig:captm_fix} can be compared to the deterministic theory.

\begin{figure}[t!]
  \centering
  \includegraphics[scale=0.42,angle=0]{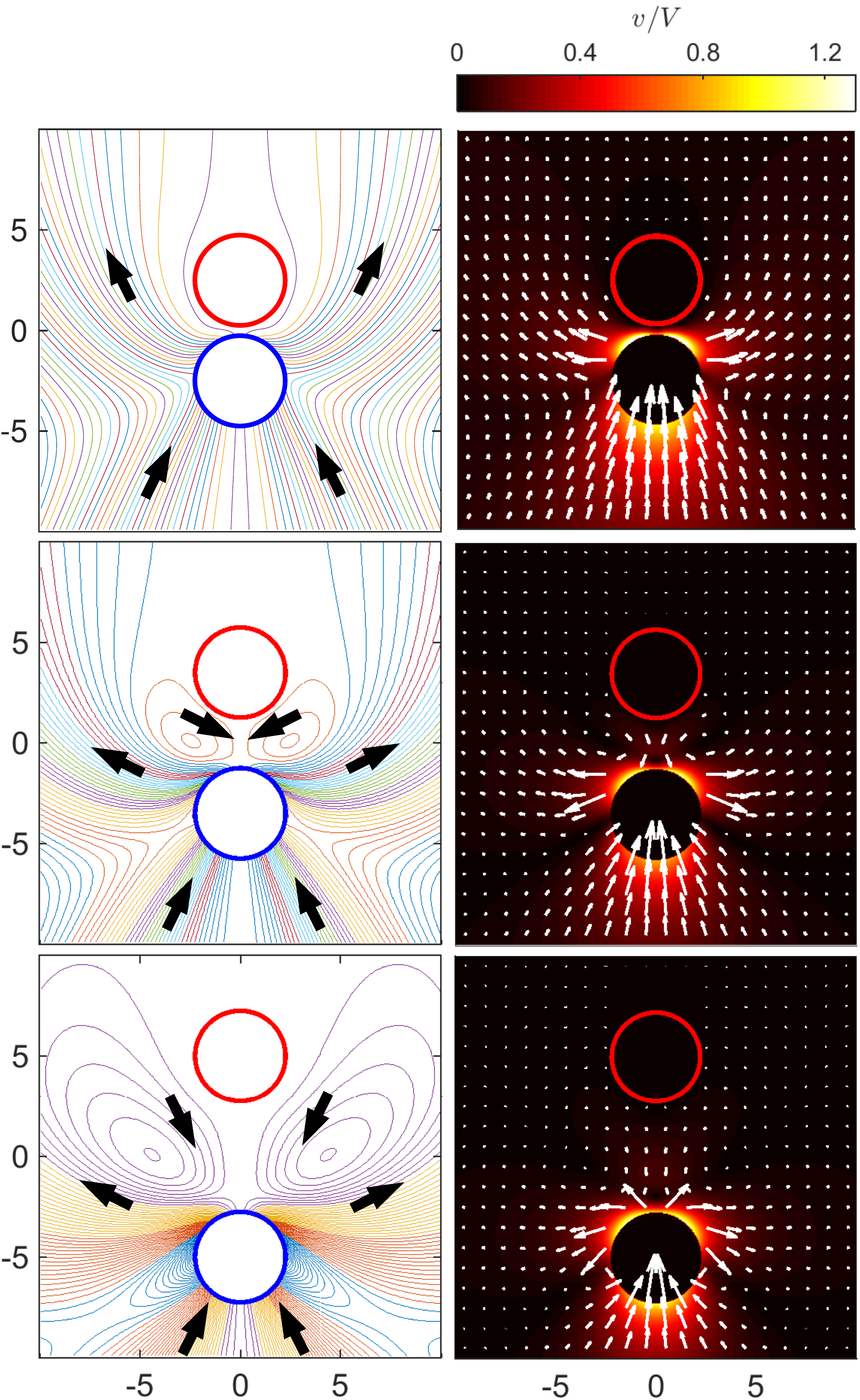}
  \caption{Streamlines and flow fields in the laboratory frame of reference.
        In the left column, streamlines are shown near the two spheres with flow directions indicated (black arrows) and, in the right column, the flow fields (white arrows) and their magnitudes (color maps),
    $v=\sqrt{v_\theta^2+v_\eta^2}$, are presented.
    The first, second, third rows are for $L/\sigma=2.5, 3.5, 5$.
    In the color maps, the magnitude of the fluid velocity $v$ is scaled by the sphere velocity $V$,
    where $V=0.053, 0.023, 0.011$ for $L/\sigma=2.5, 3.5, 5$, respectively.
    The red and blue circles indicate the $S_1$ catalytic and $S_2$
    noncatalytic spheres.
  }
  \label{fig:stream_fix}
\end{figure}
\begin{figure}[t!]
  \centering
  \includegraphics[scale=0.4,angle=0]{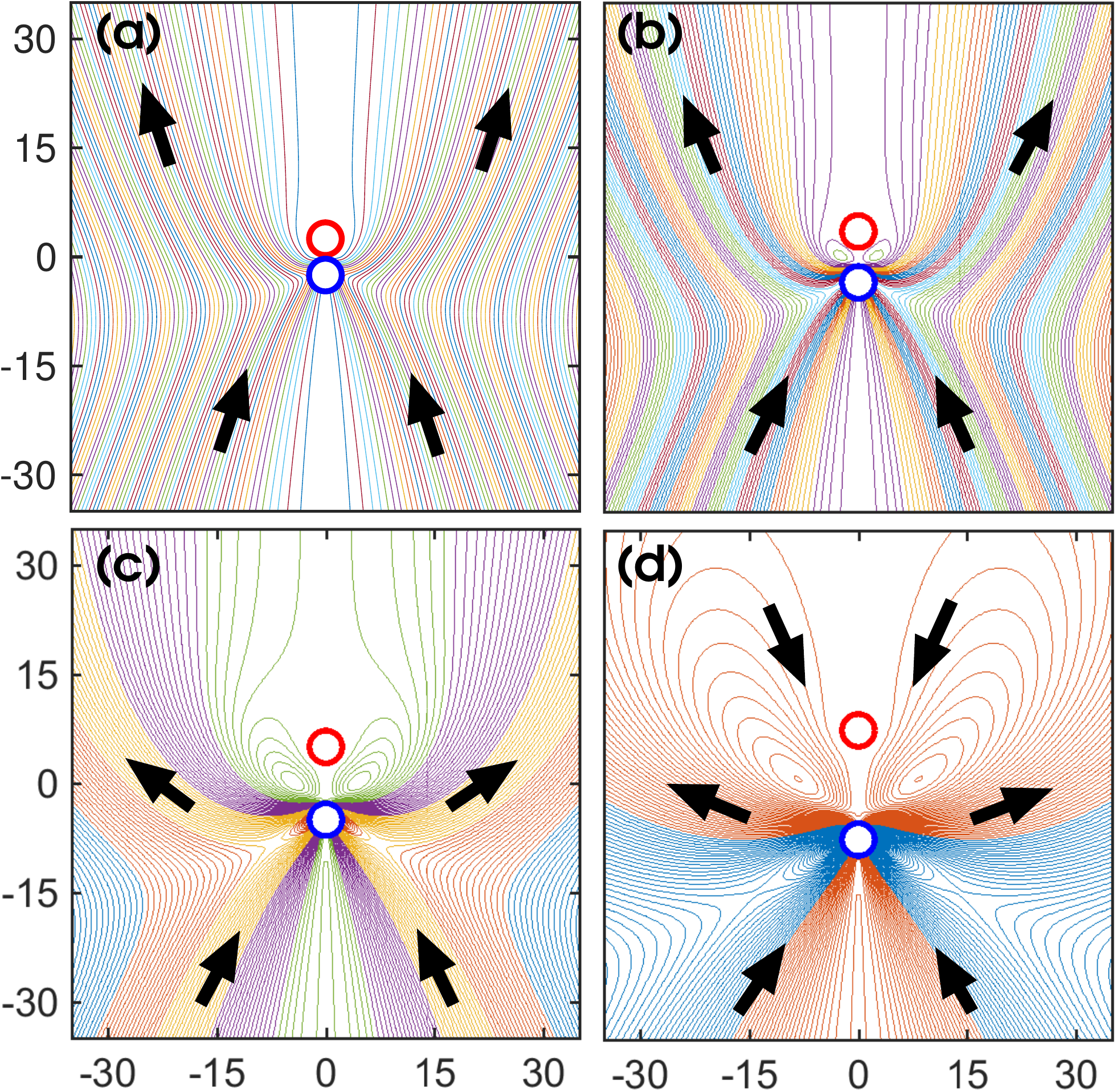}
  \caption{Far-field streamlines for various sphere separations,
    (a) $L/\sigma=2.5$ (b) $L/\sigma=3.5$ (c) $L/\sigma=5$ (d) $L/\sigma=7.5$.
      }
\label{fig:stream_far_fix}
\end{figure}
\begin{figure}[htbp!]
  \centering
  \includegraphics[scale=0.6,angle=0]{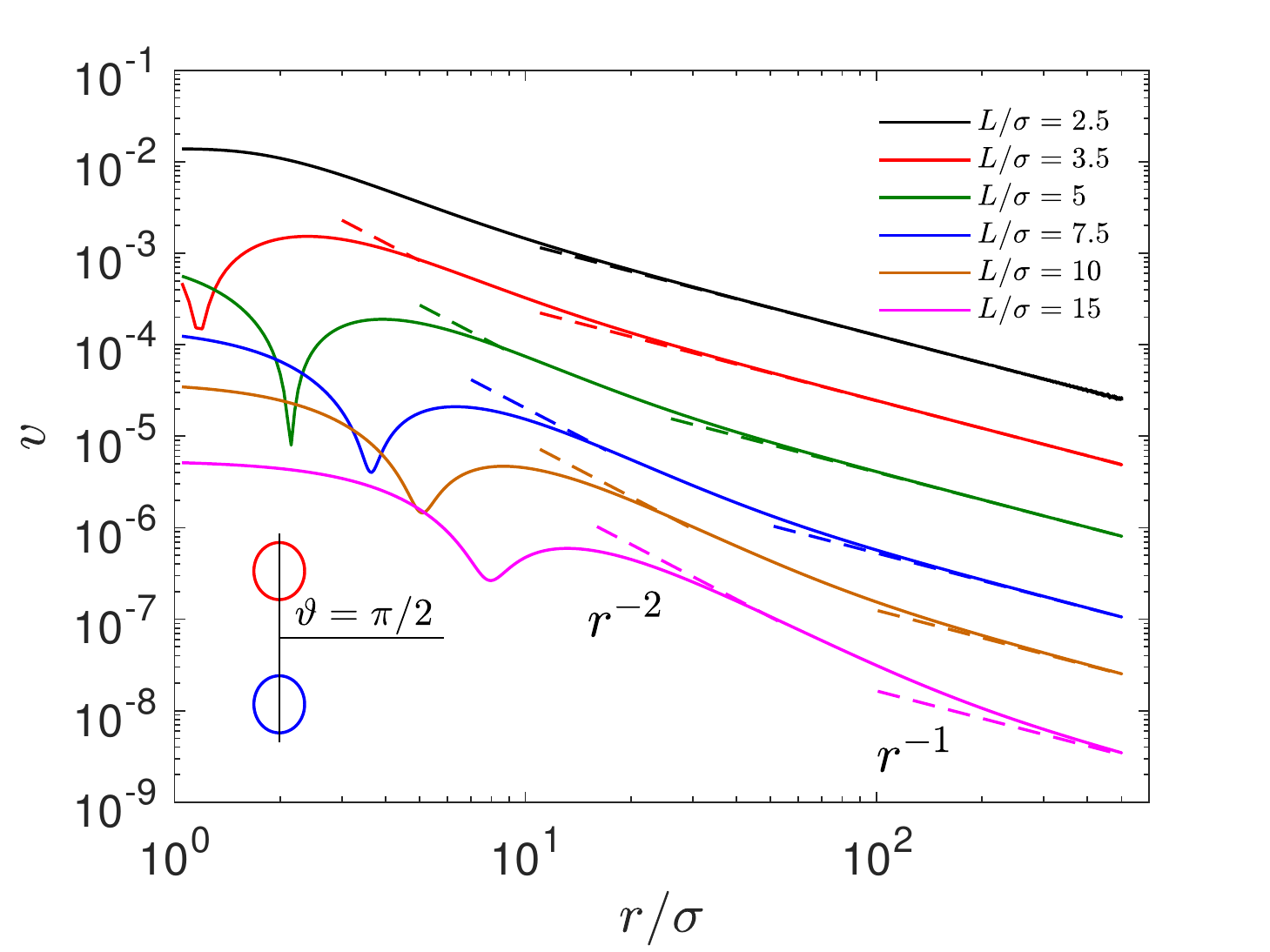}
  \caption{The magnitude of fluid velocity, $v=\sqrt{v_\theta^2+v_\eta^2}$,
    for $\vartheta=\pi/2$ as a function of distance $r$,
    where the spherical polar coordinates ($r$, $\vartheta$, $\phi$) are taken
    with a common origin in Fig.~\ref{fig:coord}.
    The black, red, green, blue, brown, magenta lines (from top to bottom)
    correspond to the separation distances, $L/\sigma=2.5$, $3.5$, $5$, $7.5$,
    $10$, $15$, respectively.
}
  \label{fig:fd_mag_fix}
\end{figure}
The capture time, $\tau$, which is defined by the time it takes the
$S_2$ sphere, initially at $L$, to reach the $S_1$
sphere, i.e., the spheres are separated by a distance equal to the sum of
their radii, $R_1+R_2$. The time $\tau$ can be calculated easily by
integrating the velocity (Eq.~(\ref{appr_vel})) to obtain the simple expression,
$\tau = (k_0+k_D)(L^3-(R_1+R_2)^3)/(2\kappa c_0k_0R_1)$.
Figure~\ref{fig:captm_fix} shows how $\tau$ varies with $L$.
The exact continuum solutions agree well with simulations, while here are
discrepancies with the approximate theory.

The concentration and fluid velocity fields vary during the capture process,
and these variations play a role in determining the details of the capture
mechanism. The $B$ species concentration fields and their gradients on the
surface of the $S_2$ sphere are shown in Fig.~\ref{fig:conc_fd}. The
concentration field decays as $1/r$ at long distances~\cite{reigh:15dimer}
but again there are discrepancies in the magnitude of the field close to the $S_2$ sphere. Such discrepancies might be expected because the dynamics in the finite-size boundary layer cannot be simply represented by the continuum   boundary conditions. It is interesting that the tangential gradient of this field on the surface corresponds very closely to that of the continuum
model. Consequently, even though the microscopic nature of the concentration fields is manifest in the boundary layer, the gradient, which determines the propulsion, is accurately given by the continuum theory. As a result many of the other observable properties are accurately given.

The velocity fields generated by the moving $S_2$ sphere present a more interesting and complex structure as a function of $L$. Figure~\ref{fig:stream_fix} shows the streamlines and flow fields in the laboratory frame of reference.
The streamlines are plotted by setting $\psi$ equal to a constant. At large separations, we see that the fluid near the head of the $S_2$ sphere (portion closest to the $S_1$ sphere) is pushed
to the lateral directions (in the $xy$ plane) with respect to the axisymmetric $z$ axis, and executes broad fluid circulation near the $S_1$ sphere. Fluid also flows towards the rear of the $S_2$ sphere.
The flow near the $S_2$ sphere shows a puller-like behavior; i.e., fluid
enters from the front and back and is expelled from the sides.~\cite{ishikawa06,stark:16}
(A pusher-like behavior can be also seen in our system if $\epsilon_B >
\epsilon_A$.)
As the two spheres approach each other ($L/\sigma \sim 3.5$) the circulating flows between and to the sides of the spheres reduce in size and disappear, leaving a  puller-like flow pattern. Near the contact distance ($L/\sigma \sim 2.5$),
the fluid is pushed from the back to the front of two spheres.

\begin{figure}
  \centering
  \includegraphics[scale=0.8,angle=0]{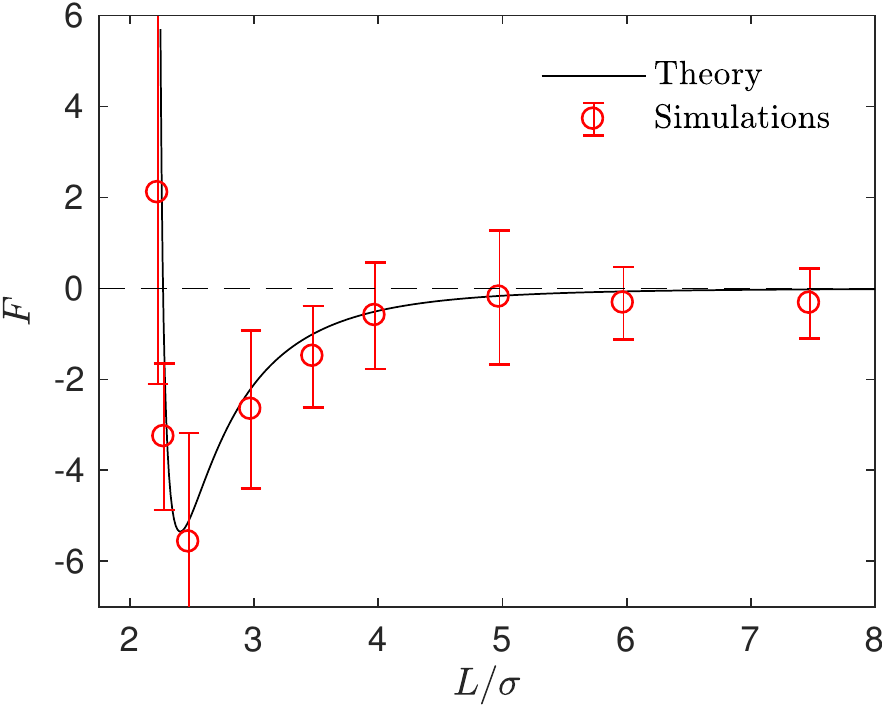}
  \caption{The force on the catalytic sphere exerted by fluid.
    The black solid line and red circles correspond to the continuum theory and simulations,
    respectively. Negative values ($-z$ direction in
    Fig.~\ref{fig:coord}) imply the force is attractive.
  }
  \label{fig:force}
\end{figure}

That the flow patterns are affected by the pinning of the catalytic sphere are clearly seen in the plots
of the far field streamlines in Fig.~\ref{fig:stream_far_fix}.
The flow near the spheres resembles that due to stresslet fields
(similar to that for $L/\sigma \sim 3.5$ and $\sim 5$ in Fig.~\ref{fig:stream_fix}), but
at distances far from the spheres (see Fig.~\ref{fig:stream_far_fix} (b) and (c)) the flow resembles a drift flow (Stokeslet).~\cite{kim:91}
When the separation between the spheres is large (Fig.~\ref{fig:stream_far_fix} (d), $L/\sigma
\sim 7.5$), the flow circulation (stresslet fields) expands to
occupy a larger portion of space, but a drift flow (Stokeslet) again appears when viewed at large distances from the spheres.
These far-field flows are characterized quantitatively by calculating
the magnitude of fluid velocity $v=\sqrt{v_\theta^2+v_\eta^2}$, where $\bm{v}
= v_\theta \hat{\bm{\theta}} + v_\eta \hat{\bm{\eta}}$,~\cite{happel73}
as shown in Fig.~\ref{fig:fd_mag_fix}.
For example, at $L/\sigma=7.5$, one sees a $1/r^2$ decay, characteristic of stresslets, for distances up
to approximately $r/\sigma \sim 20$, but eventually the flow velocity decays asymptotically as $1/r$.
As the separation distance decreases, it is notable that the flow velocity
increases, the stresslet contribution disappears, and the Stokeslet contribution increases.
The asymptotic expressions are found by introducing the spherical polar
coordinates ($r$, $\vartheta$, $\phi$) in Fig.~\ref{fig:coord},
where two coordinate systems share the origin, and
expanding the variables $\theta$ and $\eta$ in terms of $1/r$.
Then one may obtain asymptotic expressions for flow velocity up to $\mathcal{O}{(1/r^2)}$ as
\begin{align}
  &v_\theta \sim \sqrt{2} \sin \vartheta \{ 3\Omega_1\cos\vartheta /(2\xi r)
  -\Omega_2 (1-3 \cos^2 \vartheta) /r^2 \}, \nonumber \\
  &v_\eta \sim \sqrt{2} (2-3\sin^2\vartheta)\{\Omega_1/(2\xi r)
  + \Omega_1 \cos \vartheta / r^2\},
  \label{asym}
\end{align}
where $\Omega_1= \sum_{n=1}^\infty (2n+1)(a_n+c_n)$ and
$\Omega_2=\sum_{n=1}^\infty (2n+1)\{(n-1/2)b_n + (n+3/2)d_n \}$.
The details are given in the Appendix.

Since the fluid between the spheres flows from the $S_1$ to
$S_2$ spheres with a broad circulation pattern,
one may expect that the force the fluid exerts on the fixed catalytic sphere is in the same direction;
i.e., an attractive force. (If $\epsilon_B > \epsilon_A$ then the flow directions are reversed and one has a repulsive force.)
The force is give by Eq.~(\ref{eq:force_fix}) in the Appendix and is
plotted in Fig.~\ref{fig:force}, along with the simulation result.
In the microscopic simulations, the force is calculated by summing the forces on the catalytic
sphere due to all of the fluid particles.
The continuum theory and simulations agree very well.
The force is almost zero for large $L$, and becomes more negative (attractive)
as $L$ decreases, reaching its largest negative value at $L/\sigma \sim 2.5$,
near the contact distance, $L/\sigma \sim 2.25$. If $L$ decreases further, the
force take positive  (repulsive) values.

\section{Dynamics with a moving catalytic sphere}\label{sec:move}
\begin{figure}[t!]
  \centering
  \includegraphics[scale=0.57,angle=0]{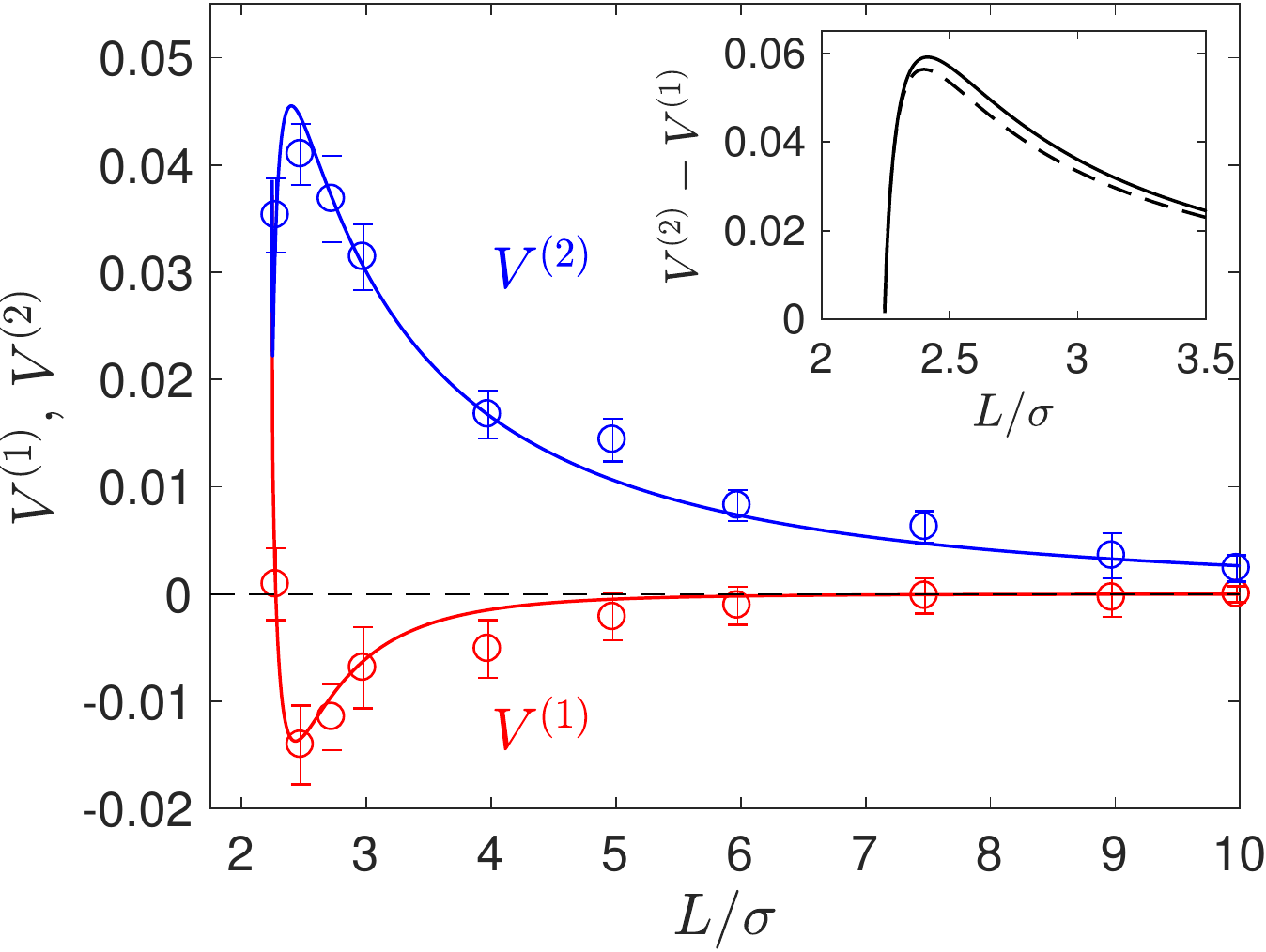}
  \caption{Plot of the velocities $V^{(1)}$ and $V^{(2)}$ of the $S_1$ and $S_2$ spheres in a force-free system.
    The solid blue and red lines denote the continuum theoretical values of $V^{(1)}$ and $V^{(2)}$, respectively, while the circles with error bars are the microscopic simulation results.
    The inset shows the velocity difference $V^{(2)}-V^{(1)}$ (solid lines) and , for comparison, the velocity of the $S_2$ sphere (dashed line) when the $S_1$ sphere is fixed in space (Eq.~\ref{mvel_fix}).
  }
  \label{fig:mvel_free}
\end{figure}
\begin{figure*}[t!]%[htbp]
  \centering
  \includegraphics[scale=0.4,angle=0]{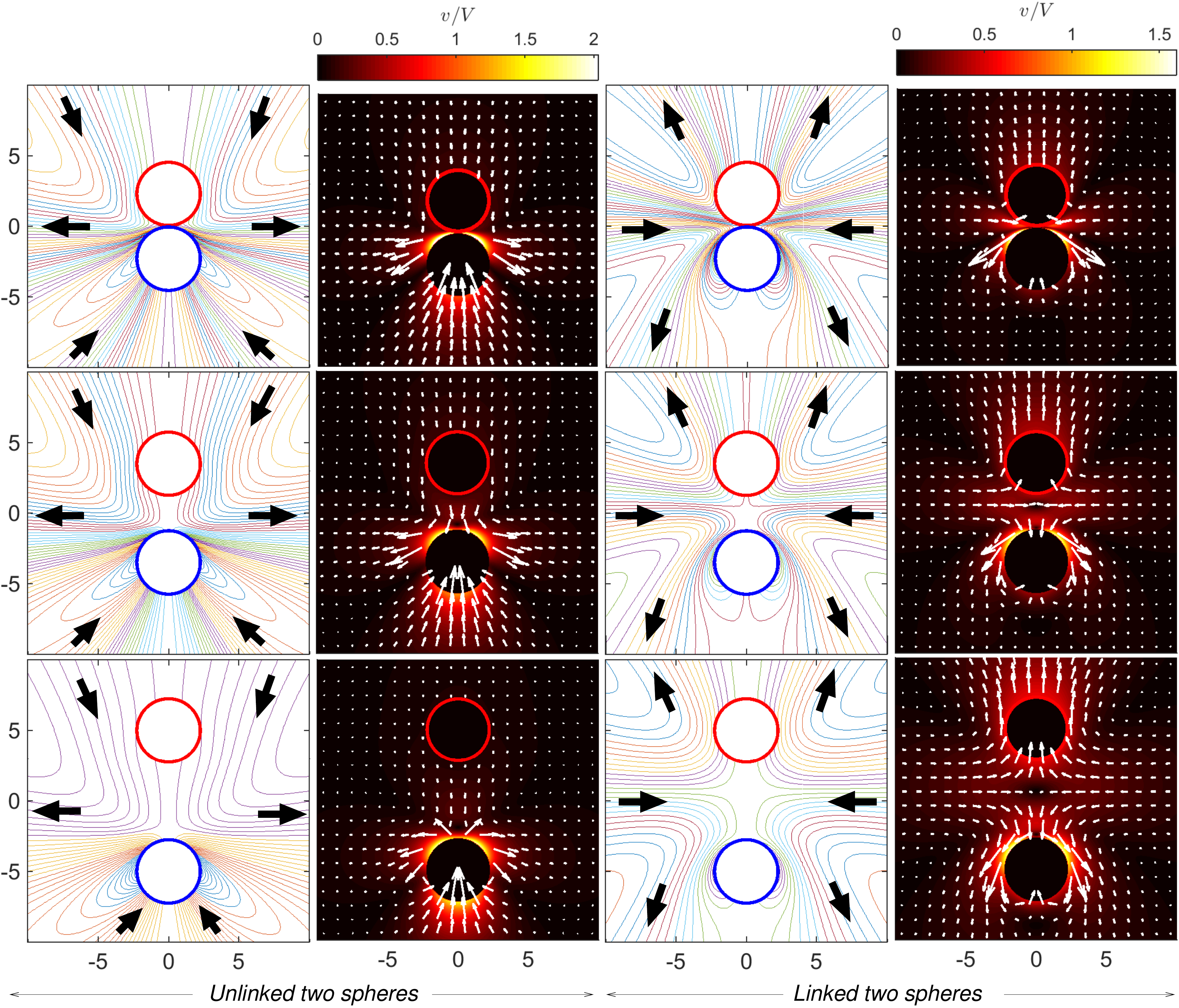}
  \caption{
    The streamlines and flow fields for the unlinked two spheres (left two
    columns) and for the linked two spheres (right two columns)
    in the laboratory frame of reference.
    The first, second, third rows correspond to the separation distances
    $L/\sigma=2.3, 3.5, 5$, respectively.
    In the color maps, the flow velocity ($v$) is scaled by the velocity of
    noncatalytic spheres ($V^{(2)}$) and dimers ($V_D$),
    where $V^{(2)}=0.039, 0.022, 0.011$ and $V_D=0.053, 0.019, 0.0084$
    in $L/\sigma=2.3, 3.5, 5$, respectively.
    The red and blue circles indicate the catalytic and noncatalytic spheres.
      }
  \label{fig:fd_comp}
\end{figure*}

We now consider the situation where both spheres are free to move. The concentration fields produced by the catalytic sphere are unchanged from the fixed-sphere case. Using the continuum theory, the velocities of both spheres can be computed from Eq.~(\ref{eq:mvel_mov}) and they are plotted in Fig.~\ref{fig:mvel_free}, along with the simulation results.
The continuum theory and microscopic simulation results are in good
agreement. Now the $S_1$ and $S_2$ spheres move towards each other, but with
different velocities as shown in the figure. The velocity of the $S_2$ sphere
is much larger than that of the $S_1$ sphere, and the velocity difference
$V^{(2)}-V^{(1)}$ is shown in the inset of the figure. For comparison, this difference
is compared with that for a fixed $S_1$ sphere, $V=V^{(1)}=0$, (dashed line in the
inset). Although the $S_2$ sphere moves by the diffusiophoretic mechanism, the
motion of the $S_1 $ sphere is induced by the fluid flow generated by the $S_2$ sphere.

Note that although the velocities of the two spheres have opposite signs (-
for $S_1$ and + for $S_2$) as they approach, the sign of the $S_1$ velocity
changes so that both sphere velocities are positive ($+z$) as the two spheres
meet to form a self-propelled sphere-dimer that moves with the $S_1$
sphere at its head (see Movie 2).~\cite{reigh:15dimer,ozin:10} In
contrast to the sphere-dimer motors previously studied that are made from
spheres with a rigid bond, this sphere-dimer motor self-assembles from
  isolated spheres to form a bound pair with a bond length that may
  fluctuate around a mean value depending on parameters used. Once
the sphere dimer is formed by self-assembly it behaves like the sphere-dimer with a fixed bond length.  Similar motion of two spheres was observed in a numerical study of a thermocapillary system consisting of a solid particle and a gas bubble.~\cite{golovin:95}

The streamlines and flow field are shown in Fig.~\ref{fig:fd_comp} (left two columns)
in the laboratory frame of reference. When $L$ is relatively large ($L/\sigma=5$), the
streamlines are roughly similar to those when the $S_1$ sphere is fixed but there is
no local fluid circulations at small distances from the spheres and no drift flow at large distances.
The fluid flow near the $S_2$ sphere exhibits a puller-like pattern and near the $S_1$ sphere fluid is simply dragged to the $S_2$ sphere. As discussed above, this difference is attributed to the contributions of Stokeslets in a forced system and these effects are pronounced at small $L$ ($L/\sigma=2.3, 3.5$).
The streamlines in a force-free system do not significantly change at small
separations, while those in a forced system are more distorted in the direction of the applied external force (Fig.~\ref{fig:stream_fix}).
The quantitative variations of streamlines and flow fields can be
seen by plotting the magnitudes of flow velocity as displayed
in Fig.~\ref{fig:fv_mag} (left panel).
The flow velocity of force-free spheres decays as a $r^{-2}$ (stresslet) in a
distance $r/\sigma \sim 5$ for various values of $L$, and
this power-law behavior remains unchanged at long distances.
However, the flow velocity in a system with sphere $S_1$ fixed exhibits a $r^{-2}$ decay for
distances $r/\sigma \sim 5$ when $L/\sigma=5$, and it shows a
$r^{-1}$ decay (Stokeslet) for $L/\sigma=2.5$, although the velocity in all
cases eventually decays a $r^{-1}$ at long distances (Fig.~\ref{fig:fd_mag_fix}).

\subsection*{Flow field comparison}
\begin{figure}[t] %[htbp]
  \centering
  \includegraphics[scale=0.45,angle=0]{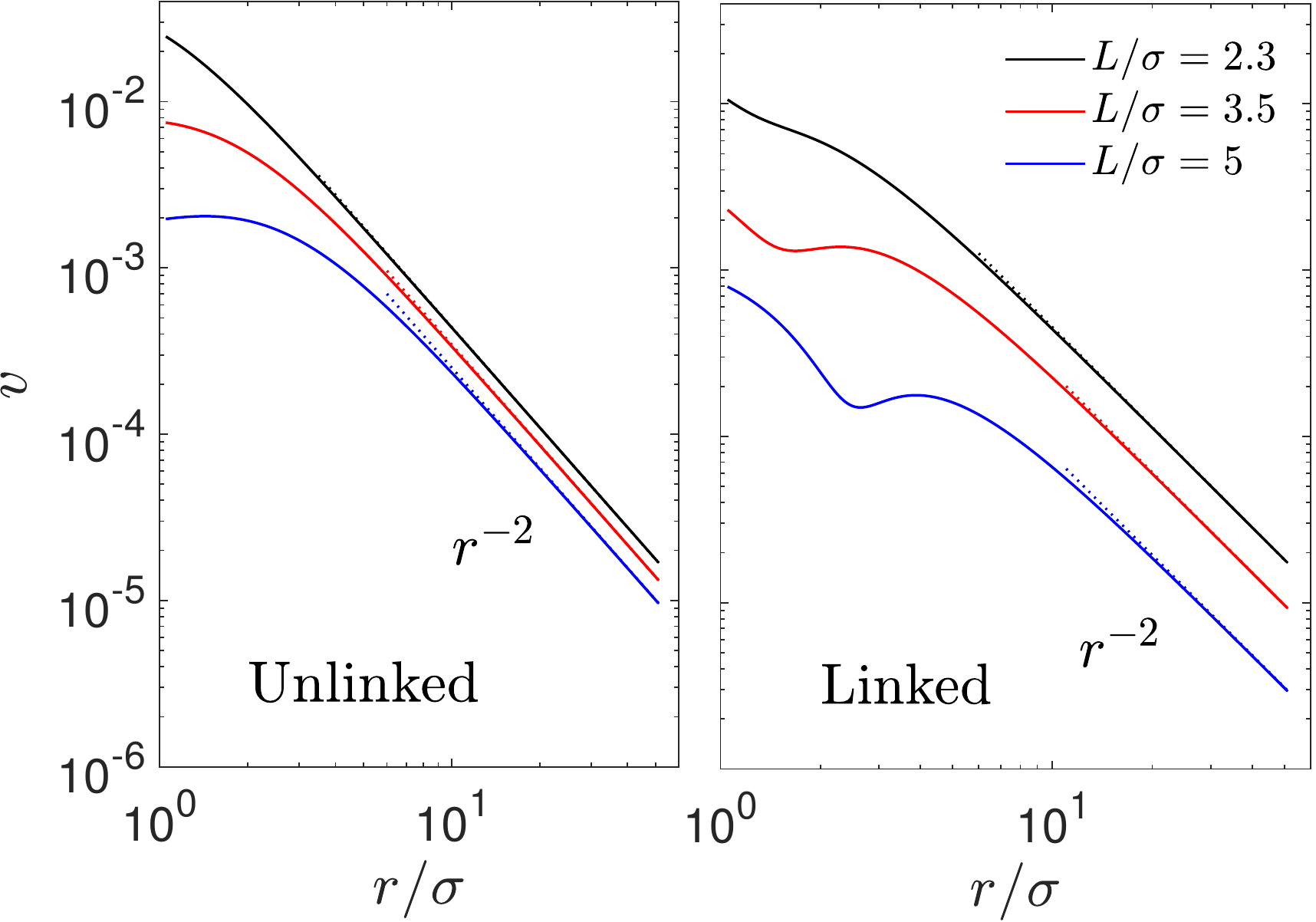}
  \caption{The magnitude of fluid velocity, $v=\sqrt{v_\theta^2+v_\eta^2}$,
    along the side direction ($\vartheta=\pi/2$) as a function of distance $r$
    for the unlinked two spheres (left) and the linked dimer (right).
    The spherical polar coordinate ($r$,$\vartheta$,$\phi$) is taken
    by setting the origin of the coordinate at the middle of two spheres
    as in Fig.~\ref{fig:coord} and Fig.~\ref{fig:fd_mag_fix}.
    The black, red, and blue lines correspond to the separation distance,
    $L/\sigma=2.3, 3.5, 5$, respectively.
}
  \label{fig:fv_mag}
\end{figure}

It is interesting to compare the properties of the flow fields for the freely
moving catalytic and noncatalytic spheres separated by a distance $L$ with those for a sphere-dimer with a rigid bond of length $L$. We refer to the spheres in the former case as {\em unlinked spheres} and those
  in the latter case as {\em linked spheres}.
We consider the unlinked spheres to be the linked when the spheres form a dimer by self-assembly.
The streamlines and flow fields just before and after the spheres self-assemble to
form a sphere-dimer motor
are shown in Fig.~\ref{fig:fd_comp} (first row). It is notable that the flow directions for the unlinked spheres (first panel in this row) are completely reversed after the spheres self-assemble to form a sphere-pair (third panel in this row), although the detailed structure of the flow field  changes near the $S_2$ sphere.
This implies that a sudden change in flow field occurs from a puller-like flow pattern to a pusher-like pattern.

These puller and pusher flow patterns remain unchanged as $L$ increases (second and third rows in Fig.~\ref{fig:fd_comp}).
The magnitudes of flow velocity for the unlinked and linked spheres are
compared quantitatively in Fig.~\ref{fig:fv_mag}. Both cases exhibit a $r^{-2}$
decay in contrast to that for a fixed $S_1$ sphere.
For small $L$ ($L/\sigma < 3.5$), the magnitudes of flow velocity for
both linked and unlinked spheres are very similar; only the flow directions have opposite signs.
The asymptotic expressions are given by Eq.~(\ref{asym}) without $\Omega_1$
terms since $\Omega_1$ is zero by the force-free condition.

\subsection*{Sphere size effects}
\begin{figure}[t] %[htbp]
  \centering
  \includegraphics[scale=0.4,angle=0]{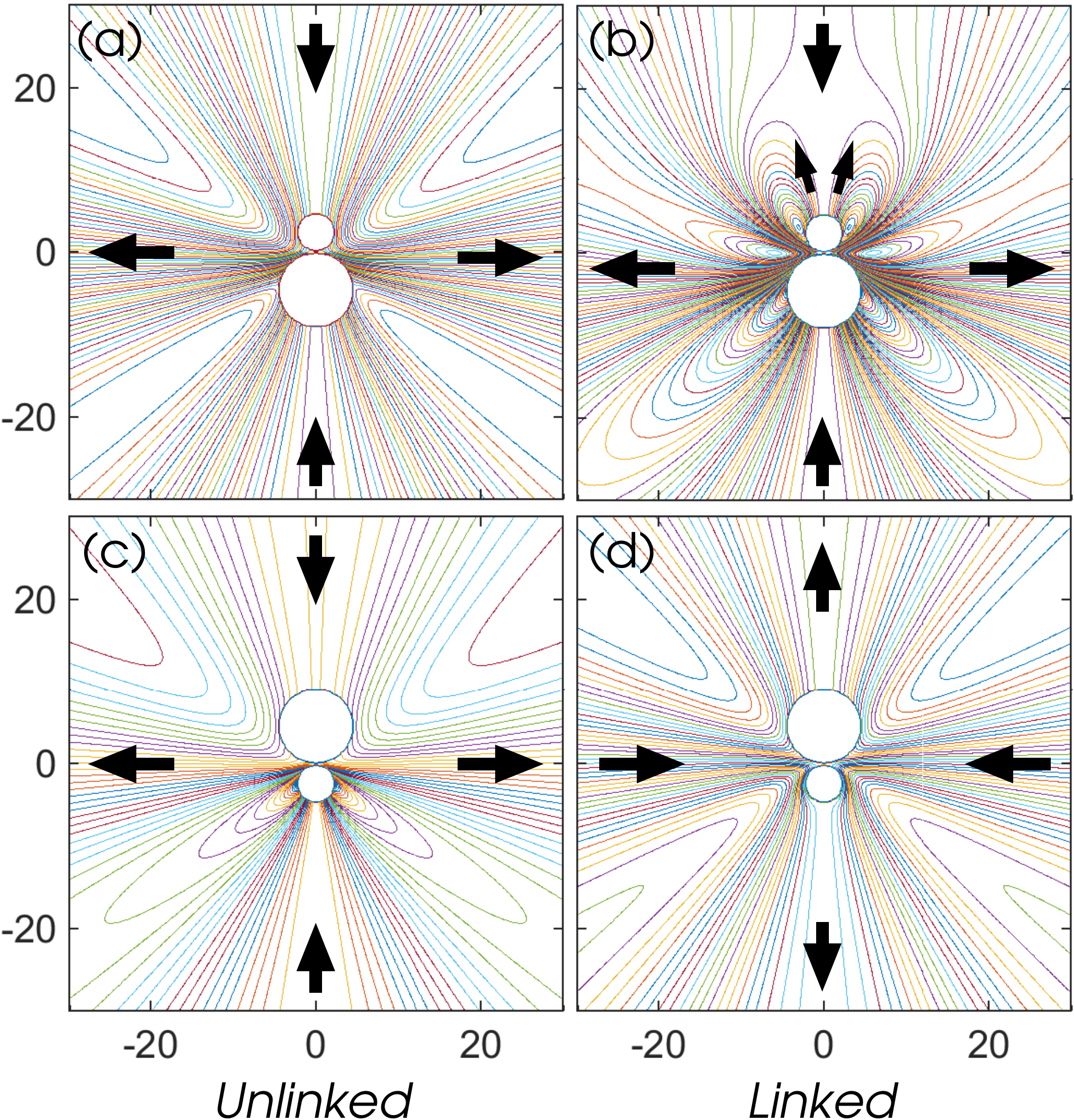}
  \caption{Streamlines before and after dimer formation for different size
    ratios of two spheres.
    The left column ((a) and (c)) shows the streamlines for unlinked
    spheres and the right ((b) and (d)) for linked spheres.
    The first row ((a) and (b)) and the second ((c) and (d))
    correspond to the size ratio between the $S_1$ and $S_2$
    spheres $R_1/R_2=0.5$ and $2$, respectively.
    The separation distances between spheres are $L/\sigma=3.5$,
    where $\sigma$ is for the small spheres, i.e. $\sigma/a=2$.
    }
  \label{fig:stream_size}
\end{figure}
Lastly, we consider how the flow fields depend on ratios of the sizes of $S_1$ and $S_2$ spheres at the moment of dimer formation.
Figure~\ref{fig:stream_size} presents the streamlines for the unlinked and linked
spheres near the contact distance, i.e. just before and after a dimer formation.
When the $S_1$  sphere is larger than the $S_2$ sphere
(Fig.~\ref{fig:stream_size} (c) and (d)),
the flow directions are completely reversed, except for local
variations near the $S_2$ sphere, similar to that for spheres of equal size: a
puller-like flow pattern changes to a pusher-like pattern. By contrast, if the
radius of the $S_1$ sphere is smaller than that of the
$S_2$ sphere (Fig.~\ref{fig:stream_size} (a) and (b)),
the character of the far-field flow does not change and is puller-like
before and after dimer formation, although the detailed structure of flow near the dimer
becomes complex and exhibits several local flow circulations, especially near the
$S_1$ sphere where fluid is pushed in the direction of its head.
It is interesting to note that two separated spheres with either size ratio are initially attracted and meet to form a
dimer, and this dimer may have one of two counter far-field flow characteristics: either a puller or pusher depending on the size ratio.

\section{Conclusions} \label{sec:conc}

Using continuum theory and particle-based simulations, a detailed study of the chemical and hydrodynamic processes that govern the dynamics of two spheres, one reactive and the other nonreactive but able to move toward high product concentrations by a diffusiophoretic mechanism, was presented in this paper. Through an analysis of the concentration and fluid flow fields the roles played by these chemical and hydrodynamic interactions could be determined. For example, when both spheres are free to move, they are attracted to each other; the nonreactive sphere moves towards the reactive sphere by diffusiophoresis while the reactive sphere is simply dragged by the flow generated by the nonreactive sphere. When the spheres are in close proximity this motion must cease; the velocity of the reactive sphere changes its sign since the nonreactive sphere now drives the pair forward by the same diffusiophoretic mechanism that operates for a sphere-dimer motor with a rigid bond. The flow field must reorganize to accommodate this change  and adopts a pusher character.

The characteristics of the flow fields depend on the sphere sizes. Two separated spheres behave as a puller, regardless of their sphere size ratio, while the sphere-dimer motor that is formed can have either puller or pusher characteristics, and this does depend on the size ratio. Consequently, it should be possible to construct self-propelled dimers with either of these flow characteristics by simply manipulating the sphere sizes. This feature may be used to aid in the understanding of the collective behavior of many-sphere systems, and to provide a route to the construction of complex self-assembled structures in
the laboratory.~\cite{palacci:13,buttinoni:13,singh:17}

The two-sphere dynamics studied in this paper may be regarded as an elementary process that contributes to the collective dynamics of mixtures of active and passive particles~\cite{balazs:17,schmidt:18,Yu:18} and sphere dimers with non-rigid bonds. The study  provides insight into the mechanisms that could lead to dynamic clusters of various types that not only move but may also fragment and reassemble. In this connection, situations not considered in this paper could be of considerable interest to investigate further. If the interactions are such that the nonreactive sphere moves to lower product concentrations, in dilute solution the two sphere will simply avoid each other. However, in more dense colloidal suspensions they will be forced to interact and lead to different active collective states, analogous to the different collective dynamics of forward and backward moving sphere dimers.~\cite{colberg:17}

\section*{Appendix}\label{appendix}
\subsection*{Continuum solution information}
The Table in this Appendix gives the definitions of functions that enter in the continuum solution.\\

\subsection*{Asymptotics of fluid velocity field}
Reminding the fluid velocity is given by the stream function as
$\bm{v}=\hat{\bm{\phi}}/\rho \times \nabla \psi$,
one gets the velocity components in $\theta$ and $\eta$, 
($v_{\theta}$, $v_{\eta}$) = 
$\{(\cosh\eta-\mu)/(\rho\xi)\}$(-$\partial \psi / \partial \eta$, $\partial
\psi / \partial \theta$) leading to
\begin{align}
  &v_{\theta} = -\frac{\sqrt{\cosh\eta-\mu}}{\xi^2\sin\theta}
  \bigg[ \sum_{n=1}^{\infty}\frac{d W_n}{d\eta} V_n \nonumber\\
  &\hspace{90pt} -\frac{3\sinh\eta}{2(\cosh\eta-\mu)}\sum_{n=1}^{\infty}W_nV_n \bigg], \nonumber\\
  &v_{\eta} = \frac{\sqrt{\cosh\eta-\mu}}{\xi^2}
  \bigg[ \sum_{n=1}^{\infty} (2n+1)W_nP_n \nonumber\\
  &\hspace{90pt} -\frac{3}{2(\cosh\eta-\mu)}\sum_{n=1}^{\infty}W_nV_n \bigg].
  \label{eq:fvel_approx}
\end{align}
From the relations between the bispherical and Cartesian coordinates
as shown in Sec.~\ref{sec:cont-theory}, one can show that
$\theta = \tan^{-1}\{ 2\xi\sqrt{x^2+y^2}/(x^2+y^2+z^2-\xi^2) \}$
and $\eta = \tanh^{-1} \{ 2\xi z / (x^2+y^2+z^2+\xi^2) \}$.
%\begin{align}
%  \theta &= \tan^{-1} \bigg( \frac{2\xi \sqrt{x^2+y^2}}{x^2+y^2+z^2-\xi^2}
%  \bigg), \nonumber \\
%  \eta &= \tanh^{-1} \bigg( \frac{2\xi z}{x^2+y^2+z^2+\xi^2} \bigg),
%\end{align}
In newly introduced spherical polar coordinates ($r$, $\vartheta$, $\phi$) in
Fig.~\ref{fig:coord}, where the origin is shared, 
the variables $\theta$ and $\eta$ in large $r$ are approximated by Taylor series 
as $(\theta, \eta) \sim (2\xi/r)(\sin\vartheta,
\cos\vartheta)+\mathcal{O}(1/r^3)$. 
%$\theta \sim (2\sin\vartheta)(\xi/r) + \mathcal{O}(1/r^3)$
%and $\eta \sim (2\cos\vartheta)(\xi/r) +\mathcal{O}(1/r^3)$.
Then all factors in Eq.~\ref{eq:fvel_approx} are expanded by Taylor series
again for large $r$ and the final forms are expressed by Eq.~\ref{asym} in the
main text.

\begin{table}[t!]
%  \squeezetable
  \small
  \caption{\label{tab1}The coefficients for the sphere velocity in
    Eq.~(\ref{mvel_fix}) and Eq.~(\ref{eq:mvel_mov}), and the fluid stream
    functions in Eq.~(\ref{eq:coeff1}) and Eq.~(\ref{eq:coeff2}).
    The coefficients $\bm{Y}^{(o)}=\{Y_n^{(1)}, Y_n^{(3)}, Y_n^{(5)}, Y_n^{(7)} \}$,
    $\Xi_n^{(+)}$, $\Gamma_n^{(+)}$ have the upper signs
    and $\bm{Y}^{(e)}=\{Y_n^{(2)}, Y_n^{(4)}, Y_n^{(6)}, Y_n^{(8)}\}$,
    $\Xi_n^{(-)}$, $\Gamma_n^{(-)}$ have the lower signs in the equations.
  }
  \begin{tabular*}{0.5\textwidth}{@{\extracolsep{\fill}}c}
    \hline
    \vbox{
      \begin{align}
        %% phi_l
        &\Phi_n =-\frac{n(n+1)}{2(2n+1)}  \{ e^{(n-\frac{1}{2})\eta_2}
        A_{n-1}-(2\cosh\eta_2)e^{(n+\frac{1}{2})\eta_2} A_n \nonumber\\
        &\hspace{30pt}+e^{(n+\frac{3}{2})\eta_2}
        A_{n+1}+e^{-(n-\frac{1}{2})\eta_2} B_{n-1} \nonumber \\ 
        &\hspace{30pt} 
        -(2\cosh\eta_2 )e^{-(n+\frac{1}{2})\eta_2}B_n +e^{-(n+\frac{3}{2})\eta_2}
        B_{n+1} \} \nonumber \\
        %% Delta_l %%%%%%%%%
        &\Delta_n = 4\sinh^2\{(n+\tfrac{1}{2})(\eta_1-\eta_2)\}
        -(2n+1)^2\sinh^2(\eta_1-\eta_2) \nonumber\\
        %%% f_l
        &f_n = \xi^2 n(n+1)/\{\sqrt{2}(2n-1)(2n+1)(2n+3)\} \nonumber\\
      %%%%%%% Y_1, Y_2 %%%%%%%%%
        &(Y_n^{(1)},Y_n^{(2)})= (2n+3)[
        \tfrac{1}{2}(2n+1)^2e^{(\eta_1-\eta_2)}\sinh(\eta_1-\eta_2)\nonumber\\
        &\hspace{50pt}
        -\tfrac{1}{2}(2n-1)(2n+1)e^{\mp(\eta_1+\eta_2)}\sinh(\eta_1-\eta_2)\nonumber\\
        &\hspace{50pt}
        +2e^{-(n+\frac{1}{2})(\eta_1-\eta_2)}\sinh\{(n+\tfrac{1}{2})(\eta_1-\eta_2)\}\nonumber\\
        &\hspace{50pt}
        +(2n-1)e^{\mp(n+\frac{1}{2})(\eta_1+\eta_2)}\sinh\{(n+\tfrac{1}{2})(\eta_1-\eta_2)\}  \nonumber\\
        &\hspace{50pt}
        -(2n+1)e^{\mp(n-\frac{1}{2})(\eta_1+\eta_2)}\sinh\{(n+\tfrac{3}{2})(\eta_1-\eta_2)\}
        ] \nonumber\\
        %%%%%%% Y_3, Y_4 %%%%%%%%%
        &(Y_n^{(3)},Y_n^{(4)})= \mp (2n+3)[
        \tfrac{1}{2}(2n+1)^2e^{(\eta_1-\eta_2)}\sinh(\eta_1-\eta_2)\nonumber\\
        &\hspace{50pt}
        -\tfrac{1}{2}(2n-1)(2n+1)e^{\mp(\eta_1+\eta_2)}\sinh(\eta_1-\eta_2)\nonumber\\
        &\hspace{50pt}
        +2e^{-(n+\frac{1}{2})(\eta_1-\eta_2)}\sinh\{(n+\tfrac{1}{2})(\eta_1-\eta_2)\}\nonumber\\
        &\hspace{50pt}
        -(2n-1)e^{\mp(n+\frac{1}{2})(\eta_1+\eta_2)}\sinh\{(n+\tfrac{1}{2})(\eta_1-\eta_2)\}  \nonumber\\
        &\hspace{50pt}
        +(2n+1)e^{\mp(n-\frac{1}{2})(\eta_1+\eta_2)}\sinh\{(n+\tfrac{3}{2})(\eta_1-\eta_2)\}
        ] \nonumber\\
        %%% Y_5, Y_6 %%%%%%%%%
        &(Y_n^{(5)},Y_n^{(6)})= -(2n-1)[
        -\tfrac{1}{2}(2n+1)^2e^{-(\eta_1-\eta_2)}\sinh(\eta_1-\eta_2)\nonumber\\
        &\hspace{50pt}
        +\tfrac{1}{2}(2n+1)(2n+3)e^{\pm(\eta_1+\eta_2)}\sinh(\eta_1-\eta_2)\nonumber\\
        &\hspace{50pt}
        +2e^{-(n+\frac{1}{2})(\eta_1-\eta_2)}\sinh\{(n+\tfrac{1}{2})(\eta_1-\eta_2)\}\nonumber\\
        &\hspace{50pt}
        -(2n+3)e^{\mp(n+\frac{1}{2})(\eta_1+\eta_2)}\sinh\{(n+\tfrac{1}{2})(\eta_1-\eta_2)\}  \nonumber\\
        &\hspace{50pt}
        +(2n+1)e^{\mp(n+\frac{3}{2})(\eta_1+\eta_2)}\sinh\{(n-\tfrac{1}{2})(\eta_1-\eta_2)\}
        ] \nonumber\\
        %%% Y_7, Y_8 %%%%%%%%%
        &(Y_n^{(7)},Y_n^{(8)})= \pm(2n-1)[
        -\tfrac{1}{2}(2n+1)^2e^{-(\eta_1-\eta_2)}\sinh(\eta_1-\eta_2)\nonumber\\
        &\hspace{50pt}
        +\tfrac{1}{2}(2n+1)(2n+3)e^{\pm(\eta_1+\eta_2)}\sinh(\eta_1-\eta_2)\nonumber\\
        &\hspace{50pt}
        +2e^{-(n+\frac{1}{2})(\eta_1-\eta_2)}\sinh\{(n+\tfrac{1}{2})(\eta_1-\eta_2)\}\nonumber\\
        &\hspace{50pt}
        +(2n+3)e^{\mp(n+\frac{1}{2})(\eta_1+\eta_2)}\sinh\{(n+\tfrac{1}{2})(\eta_1-\eta_2)\}  \nonumber\\
        &\hspace{50pt}
        -(2n+1)e^{\mp(n+\frac{3}{2})(\eta_1+\eta_2)}\sinh\{(n-\tfrac{1}{2})(\eta_1-\eta_2)\}
        ] \nonumber \\
        &z_n^{(1)} = -(2n+3)\sinh\{(n-\tfrac{1}{2})\eta_1\} \cosh\{(n+\tfrac{3}{2})(\eta_1-\eta_2)\} \nonumber\\
        &\hspace{20pt}+(2n+3)\sinh\{(n-\tfrac{1}{2})\eta_2\} \nonumber\\
        &\hspace{20pt}+(2n-1)\cosh\{(n-\tfrac{1}{2})\eta_1\}
        \sinh\{(n+\tfrac{3}{2})(\eta_1-\eta_2)\} \nonumber \\
        &z_n^{(2)} = (2n+3)\cosh\{(n-\tfrac{1}{2})\eta_1\} \cosh\{(n+\tfrac{3}{2})(\eta_1-\eta_2)\} \nonumber\\
        &\hspace{20pt}-(2n+3)\cosh\{(n-\tfrac{1}{2})\eta_2\}  \nonumber\\
        &\hspace{20pt}-(2n-1)\sinh\{(n-\tfrac{1}{2})\eta_1\}
        \sinh\{(n+\tfrac{3}{2})(\eta_1-\eta_2)\} \nonumber\\
        %%% z_l^3
        &z_n^{(3)} = (2n+3)\sinh\{(n-\tfrac{1}{2})(\eta_1-\eta_2)\} \cosh\{(n+\tfrac{3}{2})\eta_1\} \nonumber\\
        &\hspace{20pt}+(2n-1)\sinh\{(n+\tfrac{3}{2})\eta_2\} \nonumber\\
        &\hspace{20pt}-(2n-1)\cosh\{(n-\tfrac{1}{2})(\eta_1-\eta_2)\}
        \sinh\{(n+\tfrac{3}{2})\eta_1\} \nonumber
%        &z_n^{(2)} = (2n+3)\cosh(n-\tfrac{1}{2})\eta_1 \cosh(n+\tfrac{3}{2})(\eta_1-\eta_2) \nonumber\\
%        &\hspace{20pt}-(2n+3)\cosh(n-\tfrac{1}{2})\eta_2  \nonumber\\
%        &\hspace{20pt}-(2n-1)\sinh(n-\tfrac{1}{2})\eta_1
%        \sinh(n+\tfrac{3}{2})(\eta_1-\eta_2) \nonumber
      \end{align}
    }\\
    \hline
  \end{tabular*}
\end{table}

\begin{table}[t!]
  \small
%  \squeezetable
%  \caption*{Continued from Table~\ref{tab1}.}
  \begin{tabular*}{0.5\textwidth}{@{\extracolsep{\fill}}c}
    \hline
    \vbox{
      \begin{align}
        %%% z_l^1
%        &z_n^{(1)} = -(2n+3)\sinh(n-\tfrac{1}{2})\eta_1
%        \cosh(n+\tfrac{3}{2})(\eta_1-\eta_2) \nonumber\\
%        &\hspace{20pt}+(2n+3)\sinh(n-\tfrac{1}{2})\eta_2 \nonumber\\
%        &\hspace{20pt}+(2n-1)\cosh(n-\tfrac{1}{2})\eta_1
%        \sinh(n+\tfrac{3}{2})(\eta_1-\eta_2),\nonumber\\
        %%% z_l^2
%        &z_n^{(2)} = (2n+3)\cosh\{(n-\tfrac{1}{2})\eta_1\} \cosh\{(n+\tfrac{3}{2})(\eta_1-\eta_2)\} \nonumber\\
%        &\hspace{20pt}-(2n+3)\cosh\{(n-\tfrac{1}{2})\eta_2\}  \nonumber\\
%        &\hspace{20pt}-(2n-1)\sinh\{(n-\tfrac{1}{2})\eta_1\}
%        \sinh\{(n+\tfrac{3}{2})(\eta_1-\eta_2)\},\nonumber\\
        %%% z_l^3
%        &z_n^{(3)} = (2n+3)\sinh\{(n-\tfrac{1}{2})(\eta_1-\eta_2)\} \cosh\{(n+\tfrac{3}{2})\eta_1\} \nonumber\\
%        &\hspace{20pt}+(2n-1)\sinh\{(n+\tfrac{3}{2})\eta_2\} \nonumber\\
%        &\hspace{20pt}-(2n-1)\cosh\{(n-\tfrac{1}{2})(\eta_1-\eta_2)\}
%        \sinh\{(n+\tfrac{3}{2})\eta_1\}, \nonumber\\
        %%% z_l^4
        &z_n^{(4)} = -(2n+3)\sinh\{(n-\tfrac{1}{2})(\eta_1-\eta_2)\} \sinh\{(n+\tfrac{3}{2})\eta_1\} \nonumber\\
        &\hspace{20pt}-(2n-1)\cosh\{(n+\tfrac{3}{2})\eta_2\} \nonumber\\
        &\hspace{20pt}+(2n-1)\cosh\{(n-\tfrac{1}{2})(\eta_1-\eta_2)\}
        \cosh\{(n+\tfrac{3}{2})\eta_1\} \nonumber\\
        %%% Xi %%%%
        &\Xi_n^{(\pm)} = z_n^{(1)} \pm z_n^{(2)}+z_n^{(3)} \pm z_n^{(4)} \nonumber\\
        % &\Xi_n^{(2)} = z_n^{(1)}-z_n^{(2)}+z_n^{(3)}-z_n^{(4)}. \nonumber\\
        &\Gamma_n^{(\pm)} =
        2[-(2n-1)(2n+1)e^{\pm(n+\frac{3}{2})(\eta_1+\eta_2)} \nonumber\\
        &\hspace{110pt} \times \sinh\{(n-\tfrac{1}{2})(\eta_1-\eta_2)\}
        \nonumber \\
        &\hspace{30pt}
        +2(2n-1)(2n+3)e^{\pm(n+\frac{1}{2})(\eta_1+\eta_2)} \nonumber\\
        &\hspace{110pt} \times \sinh\{(n+\tfrac{1}{2})(\eta_1-\eta_2)\} \nonumber \\
        &\hspace{30pt}
        -(2n+1)(2n+3)e^{\pm(n-\frac{1}{2})(\eta_1+\eta_2)} \nonumber\\
        &\hspace{110pt} \times \sinh\{(n+\tfrac{3}{2})(\eta_1-\eta_2)\}
        ]             \nonumber\\
        &\Gamma_n^{(0)} =
        16e^{-(n+\frac{1}{2})(\eta_1-\eta_2)}\sinh\{(n+\tfrac{1}{2})(\eta_1-\eta_2)\}
        \nonumber\\
        &\hspace{30pt} +2(2n+1)[(2n+1)^2\cosh(\eta_1-\eta_2) \nonumber\\
        &\hspace{80pt} -(2n-1)(2n+3)\cosh(\eta_1+\eta_2)\nonumber\\
        &\hspace{80pt} +2(2n+1)\sinh(\eta_1-\eta_2)]\sinh(\eta_1-\eta_2)
        \nonumber\\
        &\Gamma_n^{(+)} \equiv Y_n^{(2)}-Y_n^{(4)}+Y_n^{(6)}-Y_n^{(8)}
        \nonumber\\
        &\Gamma_n^{(-)} \equiv Y_n^{(1)}+Y_n^{(3)}+Y_n^{(5)}+Y_n^{(7)}
        \nonumber\\
        &\Gamma_n^{(0)} \equiv Y_n^{(1)}-Y_n^{(3)}+Y_n^{(5)}-Y_n^{(7)}
        = Y_n^{(2)}+Y_n^{(4)}+Y_n^{(6)}+Y_n^{(8)}  \nonumber
        %% Xi_l
%        &\Xi_n = z_n^{(1)}-z_n^{(2)}+z_n^{(3)}-z_n^{(4)}. \nonumber\\
%        &\Omega_n = \bar{a}_n - \bar{b}_n + \bar{c}_n - \bar{d}_n \nonumber
      \end{align}
    }\\
    \hline
  \end{tabular*}
\end{table}

\section*{Acknowledgements}
S. Y. Reigh greatly thanks S. Dietrich for his support and acknowledges helpful discussion with P. Fischer
in Max-Planck-Institute for Intelligent Systems. The research of RK was supported in part by a grant from the Natural Sciences and Engineering Research Council of Canada. The computational work was carried out at the HPC facility in IISER Bhopal, India.

%\begin{acknowledgments}
%\end{acknowledgments}

%\bibliography{chemo}
\providecommand*{\mcitethebibliography}{\thebibliography}
\csname @ifundefined\endcsname{endmcitethebibliography}
{\let\endmcitethebibliography\endthebibliography}{}

%\bibliography{fluid}

\begin{mcitethebibliography}{53}
\bibitem[foo()]{footnote:back}
If the reactive sphere is subject to different interaction potentials and is
  diffusiophoretically active, the effects on its dynamics will be small since
  the concentration field in its vicinity is approximately spherically
  symmetric; however, at very short distances the presence of the nonreactive
  sphere can lead to asymmetry in the concentration field.\relax
\mciteBstWouldAddEndPunctfalse
\mciteSetBstMidEndSepPunct{\mcitedefaultmidpunct}
{}{\mcitedefaultseppunct}\relax
\EndOfBibitem
\end{mcitethebibliography}

\end{document}